\documentclass{article}
\usepackage{transparent}
\usepackage{amssymb}
\usepackage{tabularx}
\usepackage{multirow}
\usepackage{bm}
\usepackage{colortbl}
\usepackage{subfigure}
\definecolor{ForestGreen}{RGB}{34,139,34}
\definecolor{LightCyan}{rgb}{0.88,1,1}

\newcommand{\Th}[1]{\textsc{#1}}

\usepackage[english]{babel}
\usepackage{float}
\usepackage{booktabs}

\usepackage{graphicx} 
\usepackage{dcolumn} 
\usepackage{bm} 
\usepackage{cite}
\usepackage{amsthm} 
\usepackage{amsmath}
\include{cip-v3-pack-and-defined}
\usepackage{comment}
\graphicspath{{figures/}}

\usepackage[letterpaper,top=2cm,bottom=2cm,left=3cm,right=3cm,marginparwidth=1.75cm]{geometry}

\usepackage{amsmath}
\usepackage{graphicx}
\usepackage[colorlinks=true, allcolors=blue]{hyperref}
\usepackage{authblk}
\usepackage{titlesec}

\title{Hard negative sampling in hyperedge prediction}
\author[a]{Zhenyu Deng}
\author[a,$\ast$]{Tao Zhou}
\author[a,$\ast$]{Yilin Bi}
\affil[a]{CompleX Lab, School of Computer Science and Engineering, University of Electronic Science and Technology of China, Chengdu 610054, China}
\affil[$\ast$]{Corresponding author(s). E-mail(s): \href{zhutou@ustc.edu}{zhutou@ustc.edu}, \href{betayilin@gmail.com}{betayilin@gmail.com}}

\begin{document}
\maketitle

\begin{abstract}
Hypergraph, which allows each hyperedge to encompass an arbitrary number of nodes, is a powerful tool for modeling multi-entity interactions. Hyperedge prediction is a fundamental task that aims to predict future hyperedges or identify existent but unobserved hyperedges based on those observed. In link prediction for simple graphs, most observed links are treated as positive samples, while all unobserved links are considered as negative samples. However, this full-sampling strategy is impractical for hyperedge prediction, due to the number of unobserved hyperedges in a hypergraph significantly exceeds the number of observed ones. Therefore, one has to utilize some negative sampling methods to generate negative samples, ensuring their quantity is comparable to that of positive samples. In current hyperedge prediction, randomly selecting negative samples is a routine practice. But through experimental analysis, we discover a critical limitation of random selecting that the generated negative samples are too easily distinguishable from positive samples. This leads to premature convergence of the model and reduces the accuracy of prediction. To overcome this issue, we propose a novel method to generate negative samples, named as hard negative sampling (HNS). Unlike traditional methods that construct negative hyperedges by selecting node sets from the original hypergraph, HNS directly synthesizes negative samples in the hyperedge embedding space, thereby generating more challenging and informative negative samples. Our results demonstrate that HNS significantly enhances both accuracy and robustness of the prediction. Moreover, as a plug-and-play technique, HNS can be easily applied in the training of various hyperedge prediction models based on representation learning.

\end{abstract}


\section{Introduction}
Graphs (\textit{i.e.}, networks) have been widely used to model collections of interacting entities \cite{chen2003, watts2011, barabasial2016, newman2018networks}, demonstrating immense potential in various domains such as social networks \cite{ejnewman2002random, jackson1996strategic, ejnewman2001structure}, transportation networks \cite{kurant2005layered, ding2019application, an2014synchronization}, and ecological networks \cite{yuan2021climate, proulx2005network, montoya2006ecological, deng2012molecular}. However, simple graphs can only capture pairwise interactions, neglecting the higher-order interactions that are prevalent in both nature and human society. Examples of such higher-order interactions include collaborations among multiple scientists, physical contacts among multiple individuals in a confined space, and groups of proteins that form functional complexes \cite{battiston2020networks, battiston2021physics, mulas2022graphs, bianconi2021higher-order, bick2023what, feng2020hypergraph, vasilyeva2021multilayer}. To overcome this limitation, the concept of hypergraphs has emerged \cite{berge1984}. Hypergraphs are a generalization of simple graphs, where a hyperedge can contain an arbitrary number of nodes, thus naturally capturing multi-entity interactions \cite{battiston2020networks, battiston2021physics, bianconi2021higher-order}.

In network science, the observed links often represent only a subset of all existent ones. This phenomenon is particularly pronounced in social networks \cite{kossinets2006} and biological networks \cite{guimera2009, devkota2020}. For instance, protein-protein interactions and gene regulatory relationships account for only a small fraction of all interactions that truly exist. Therefore, a fundamental challenge is how to predict existent but unobserved links based on the observed ones. This problem, known as link prediction, is one of the core topics in network science \cite{lü2011link, martnez2016survey, kumar2020, zhou2021progresses, carlo2023}. Similar to link prediction, hyperedge prediction aims to infer unobserved yet existent hyperedges based on the observed ones, which can be regarded as a generalization of link prediction to hypergraphs \cite{chen2024survey}. As a fundamental task in the study of hypergraph, hyperedge prediction has emerged as one of the most active areas, offering significant value in a wide range of real-world applications \cite{bairey2016high-order, klamt2009hypergraphs, nguyen2024central-smoothing, yu2011higher-order}. A common assumption in hyperedge prediction is that nodes with similar features tend to appear in the same hyperedge. Existing studies generally adopt two approaches to quantify the similarity of nodes in hypergraphs: (i) Extending methods to characterize the similarity between a pair of nodes in simple graphs to hypergraphs \cite{zhang2018link,kumar2020hpra,sharma2020c3mm}, and (ii) Learning a high-dimensional embedding space where each node is represented as a point. The similarity between two nodes is then measured by their distance in space, with a shorter distance typically indicating stronger similarity \cite{dong2020hnhn, yadati2020nhp, hwang2022ahp, li2023interpretable, behrouz2023cat-walk}.

To better represent the information of nodes in a high-dimensional embedding space, it is necessary to ensure that the connected nodes are placed close together, while the unconnected nodes are pushed farther apart. As a result, when training embedding-based models, unobserved links must be treated as negative samples. Unfortunately, we do not know which node pairs are truly unconnected, since some links, despite being unobserved, may actually exist. This uncertainty is precisely the motivation behind the link prediction. Due to the extreme sparsity of real-world networks \cite{genio2011, goswami2018}, the proportion of existent but unobserved links among all missing links is negligible. In fact, for a network with $m$ nodes, the number of existent links (in simple network) or hyperedges (in hypergraph) typically scales on the order of $O(m)$. However, the total number of possible links in a simple network is $O(m^{2})$, while in a hypergraph, the number of possible hyperedges is exponentially larger, say $O(2^{m})$. Given this, for simple networks, particularly when $m$ is relatively small, full-sampling can be employed to construct negative samples, namely to treat all unobserved links as negative candidates \cite{zhou2009predicting, hajimoradlou2022stay}. However, this method is completely infeasible for hypergraphs, as the number of unobserved hyperedges is about slightly less than $O(2^{m})$, which is a ridiculously large number. To make computation feasible, a widely adopted approach is to perform random sampling from the set of unobserved hyperedges \cite{patil2020negative, hwang2022ahp, yadati2020nhp}, selecting only a small fraction as negative samples.

Link prediction, as well as hyperedge prediction, can be fundamentally viewed as a binary classification problem \cite{hastie2009}. In such problem, once the positive samples are given, the selection of negative samples plays a crucial role in model training and classification accuracy. Generally, if negative samples are difficult to distinguish from positive samples, the trained model tends to exhibit stronger classification capability. Conversely, if the negative samples are easily distinguishable from positive ones, even if the model achieves perfect separation, it may remain mediocre and fail when confronted with more ambiguous real-world samples, ultimately reducing its prediction accuracy. As a result, improving the quality of sampling, such as by selecting more challenging positive and negative samples, has been a central focus in the field of binary classification. Well-known algorithms like AdaBoost \cite{freund1997} and Gradient Boosting \cite{friedman2001} are primarily designed to iteratively adjust the weights of the sample in loss function. This process emphasizes hard-to-classify samples, thereby enhancing the discriminability of the model. In a word, the selection of negative samples can significantly impact model performance. Unfortunately, existing studies on hyperedge prediction have largely ignored this issue. Of course, if negative samples obtained by random sampling are already highly indistinguishable from positive samples, researchers may comfortably adopt this straightforward and easy-to-implement method, reducing the practical necessity of exploring alternative negative sampling strategies. Based on these considerations, the first core question this study aims to address is: \textbf{In hyperedge prediction, are the negative samples obtained by random negative sampling easily distinguishable from the positive samples?}

\begin{figure}[ht]
	\centering
	\includegraphics[width=0.8\textwidth]{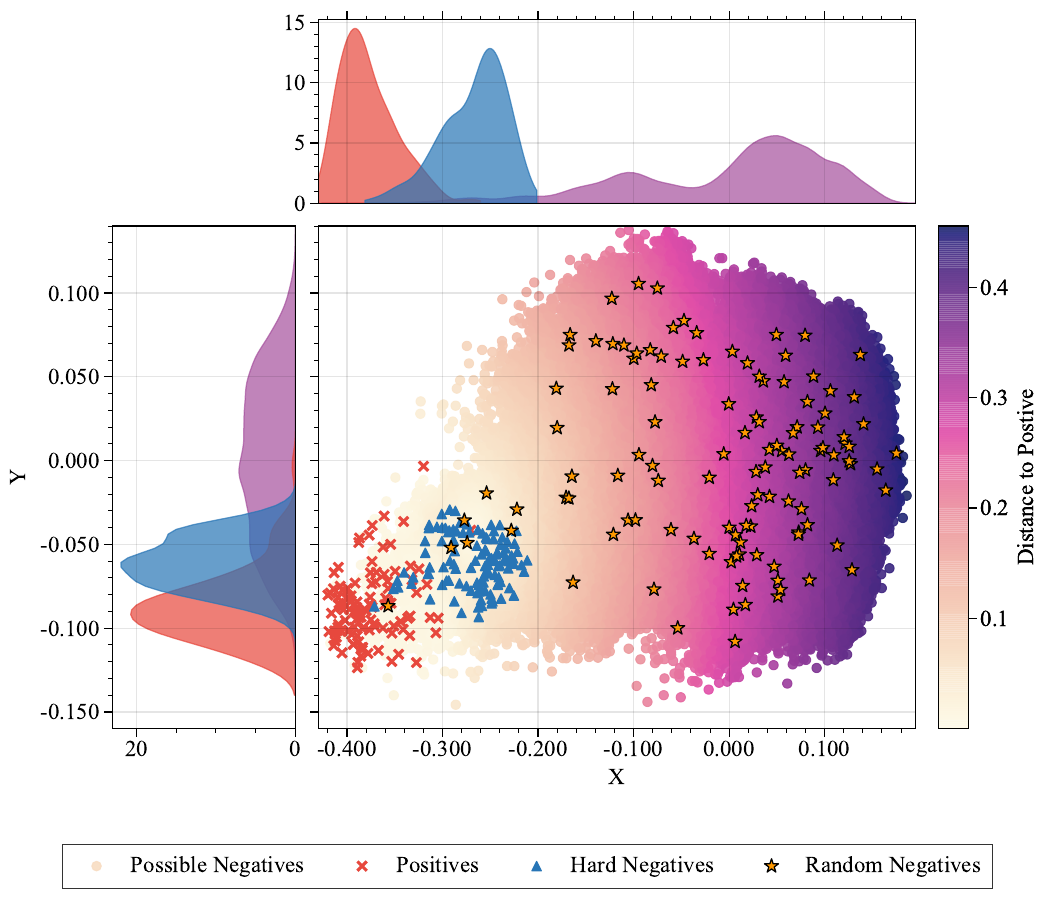}
	\caption{\textbf{Visualization of hyperedge embeddings in reduced dimensionality}. This figure displays distributions of four different types of samples for the \text{Email-Enron} dataset, projected into $2D$ space using high-dimensional embeddings obtained through self-supervised learning: (1) Possible Negatives: All potential negative samples with orders ranging from $3$ to $5$; (2) Positives: The set of positive samples; (3) Random Negatives: Negative samples randomly selected from all possible neighbors; and (4) Hard Negatives: Negative samples obtained by the HNS method.}
	\label{fig:embedding_visual}
\end{figure}

To empirically investigate this question, we extracted a subgraph from a real-world hypergraph Email-Enron \cite{Benson-2018-simplicial}, consisting of the 30 nodes with the largest hyperdegrees (a node’s hyperdegree is defined as the number of hyperedges it participates in). The positive samples are the hyperedges in this subgraph, while the candidate set of negative samples consists of all possible hyperedges between $3$-order and $5$-order that are not included in the positive samples. We applied the node2vec algorithm \cite{grover2016node2vec} to generate node embeddings, aggregated them into hyperedge embeddings, and finally performed dimensionality reduction for visualization. The results are shown in Figure \ref{fig:embedding_visual}. As illustrated in Figure \ref{fig:embedding_visual}, within the space covered by all possible negative sample candidates, all positive samples (red cross markers) cluster within a very small region in the lower-left corner. Meanwhile, when a set of negative samples equal in size to the positive samples is drawn randomly (black star markers), they are distributed far away from the positive samples, making them trivially separable. These easily distinguishable negative samples provide less useful information during model training, and using such negative samples is likely to result in incorrect classification criteria. When we applied the same approach to other hypergraphs, we obtained results that are consistent with those presented in Figure \ref{fig:embedding_visual} (please see \ref{AppendixA} in details). Consequently, we can draw a valuable conclusion, which directly answers our first question: In hyperedge prediction, the randomly selected negative samples are easily distinguishable from positive samples, limiting their usefulness in training high-performance classification models.

Closely related to the first question, the second core question of this study is: \textbf{Does there exist a method to generate negative samples that are challenging yet distinguishable from positive samples}? For hyperedge prediction, there are three main challenges: (1) Graph is non-Euclidean, making it impossible to directly quantify the distances between hyperedges; (2) The discrete nature of graph indicates that when we choose unobserved hyperedges as negative samples, there is no method to continuously adjust the distance between negative and positive samples and the positive samples; (3) The exponentially large pool of unobserved hyperedges imposes prohibitive computational costs for identifying negatives close to positives. Beyond completely random selection from unobserved hyperedges, researchers have proposed heuristic algorithms like motif negative sampling (MNS) \cite{patil2020negative} and clique negative sampling (CNS) \cite{patil2020negative} that generate negatives by replacing some nodes in positive samples. Both these heuristics and the random sampling method implicitly follow an axiomatic rule that all negatives must originate from unobserved hyperedges. We argue that this seemingly natural rule actually imposes unnecessary constraints since embedding-based methods represent each hyperedge as a point in an Euclidean space \cite{heath1956}, and thus we can measure distance and create negative samples within this continuous space. Building on this insight, we propose a novel method to generate negative samples that injects positive sample information to synthesize challenging negatives. Unlike conventional methods selecting negatives directly from unobserved hyperedges, our innovation lies in generating hard-to-distinguish negatives within the hyperedge embedding space. Specifically, we first learn continuous high-dimensional embeddings for hyperedges, then compute similarity scores between positive and negative samples based on these embeddings, and finally synthesize negative samples by injecting positive sample information. This approach not only offers greater flexibility than traditional sampling methods but also enables precise control over classification difficulty (\textit{i.e.}, the shorter the average distance between positive and negative samples, the more challenging to classify the samples accurately). Consequently, we term this method hard negative sampling (HNS). As illustrated in Figure \ref{fig:embedding_visual}, the negative samples produced by HNS (blue nodes) are notably closer to the positive samples than those obtained via random sampling (star-shaped nodes). Furthermore, the findings from other real-world hypergraphs align with those presented in Figure \ref{fig:embedding_visual} (please see \ref{AppendixA} for more details). Therefore, regarding the second core question, our answer is that such a method does indeed exist.

After obtaining affirmative answers to the previous two questions, we naturally arrive at the third core question: \textbf{Can utilizing hard-to-distinguish negative samples significantly improve model's prediction accuracy}? Intuitively, negatives closer to the boundaries of classification (\textit{i.e.}, more challenging samples) should contribute more to model training \cite{yang2024does}. However, this hypothesis has not yet been thoroughly analyzed in hyperedge prediction. In this paper, we conduct a comparative analysis of prediction accuracy across different negative sampling methods using seven real-world hypergraphs and four state-of-the-art graph embedding-based hyperedge prediction algorithms. The results show that the prediction accuracy achieved by HNS is significantly higher than that obtained through random negative sampling and other current state-of-the-art heuristic methods. Therefore, our answer to the third core question is also YES.

Our proposed method is plug-and-play, which means it is not constrained by the implementation details of other parts of the model and does not affect the usability of other components. Therefore, this approach is applicable to any method that involves representing links in a graph as vectors. Furthermore, our proposed novel method of “generating hard negatives through fusion of positive samples and easy negatives in the embedding space” can be extended to deal with node-level and subgraph-level negative sampling tasks. We believe that this methodology will advance graph analysis and mining research.

\section{Preliminaries}

\subsection{Problem Description}
We formally define a hypergraph $\mathcal{G}$ as a binary tuple $(\mathcal{V}, \mathcal{E})$, where $\mathcal{V} = \{v_1, v_2, \dots, v_m\}$ denotes the node set and $\mathcal{E} = \{e_1, e_2, \dots, e_n\}$ represents the hyperedge set \cite{berge1984, bretto2013hypergraphs}. Node features are encoded in a matrix $\mathbf{X} \in \mathbb{R}^{m \times d}$, where each row vector $\mathbf{x}_i \in \mathbb{R}^d$ corresponds to the intrinsic $d$-dimensional features of node $v_i$. Unlike simple graphs where a link connects only two nodes, a hyperedge $e_j \subseteq \mathcal{V}$ can be an arbitrary non-empty subset of nodes, enabling the representation of higher-order interactions among multiple entities. The structural information of the hypergraph is captured by an incidence matrix $\mathbf{H} \in \{0, 1\}^{m \times n}$, where the element $h_{ij}$ is set to $1$ if and only if $v_i \in e_j$; conversely, if $v_i$ does not belong to $e_j$, then $h_{ij}$ is set to $0$. The degree $d_i$ of node $v_i \in \mathcal{V}$, representing the number of hyperedges containing it, is $d_{i} = \sum_{j=1}^{n} h_{ij}$. Correspondingly, the order $c_j$ of hyperedge $e_j \in \mathcal{E}$, indicating the number of nodes it contains, is $c_{j} = \sum_{i=1}^{m} h_{ij}$.

Hyperedge prediction aims to infer unobserved or future hyperedges based on the observed hyperedge set $\mathcal{E}$ and node features $\mathbf{X}$ (generally speaking, the former is essential, while the latter may or may not be present). Let the target hyperedge set (\textit{i.e.}, unobserved or future hyperedges) be $\mathcal{E}'$ such that $\mathcal{E}' \cap \mathcal{E} = \emptyset$. This task can be formalized as a binary classification problem: Given a hypergraph $\mathcal{G}(\mathcal{V}, \mathcal{E})$ and node features $\mathbf{X} \in \mathbb{R}^{m \times d}$, for any candidate hyperedge $e \notin \mathcal{E}$, the objective is to predict whether $e$ belongs to $\mathcal{E}'$.

\subsection{Pipeline of Hyperedge Prediction}

\begin{figure}[ht]
	\centering
	\includegraphics[width=0.95\linewidth]{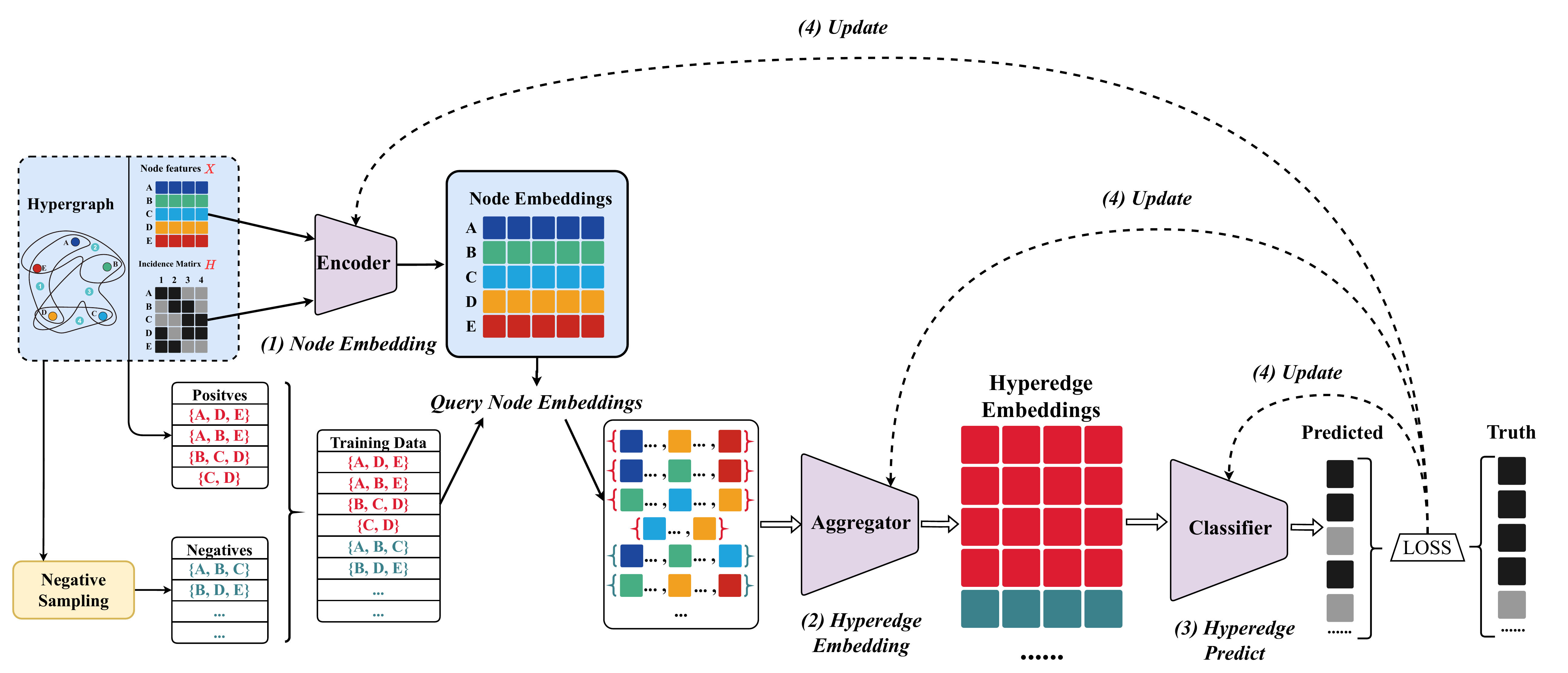}
	\caption{\textbf{The flowchart of a learning-based hyperedge prediction framework}. This framework consists of four sequential stages: (1) Node Embedding: Using hypergraph neural networks to embed nodes; (2) Hyperedge Embedding: Generating hyperedge embeddings through neighborhood aggregation; (3) Prediction: Classifying hyperedge embeddings to predict the existence of hyperedges; and (4) Loss Calculation and Model Update: Computing the loss between predicted values and true labels, and updating model parameters accordingly.}
	\label{fig:HELP_Pipeline}
\end{figure}

The embedding-based hyperedge prediction framework comprises three core components \cite{yadati2020nhp,hwang2022ahp,li2023interpretable}: An encoder, an aggregator, and a classifier. The overall architecture is illustrated in Figure \ref{fig:HELP_Pipeline}. The encoder maps nodes into a high-dimensional vector space. These embeddings capture both \textbf{attribute similarity} (reflecting intrinsic node properties) and \textbf{structural similarity} (reflecting topological closeness), forming the foundation for subsequent hyperedge prediction. The aggregator integrates the embeddings of all nodes involved in a candidate hyperedge to generate a unified hyperedge embedding. Finally, based on the aggregated embedding, the classifier achieves a likelihood score that determines whether the candidate hyperedge belongs to the target set $\mathcal{E}'$. Next, we elaborate on the implementation details of each component.

The encoder aims to capture the similarity between nodes and represent nodes as high-dimensional embedding vectors. Node similarity can be categorized into two types: Attribute similarity and structural similarity. Attribute similarity depends on the intrinsic attributes of nodes, such as occupation and interests of an individual in a social network, which are contained in the node feature matrix $\mathbf{X}$. Structural similarity depends on the roles that nodes play within the network topology, which is captured by the incidence matrix $\mathbf{H}$. The output of the node encoder is the embedding vectors of all nodes. Formally, the encoder can be defined as:
\begin{equation}
	\mathbf{V} = \operatorname{Encoder}(\mathbf{H}, \mathbf{X}),
\end{equation}
where $\mathbf{V} \in \mathbb{R}^{m \times h}$ is the representation matrix of all node, and $h$ is the dimensionality of the embedding vectors. There are multiple choices for the encoder, it can be manually constructed based on prior knowledge, such as SEHP \cite{li2023interpretable}, or it can automatically learn from data using hypergraph neural networks, such as HGNN \cite{feng2019hypergraph}, HGNNP \cite{gao2023hgnn+}, HNHN \cite{dong2020hnhn} and NHP \cite{yadati2020nhp}.

For any candidate hyperedge $e_q=\{ v_{q_1}, v_{q_2}, \dots, v_{q_{c_q}} \}$, the goal of the aggregator is to take these $c_q$ $d$-dimensional embedding vectors as input and output a single $d$-dimensional vector as the embedding vector for $e_q$. Formally, the aggregator can be defined by:
\begin{equation}
    \mathbf{e}_q = \operatorname{Aggregator}(\mathcal{V}_q),
\end{equation}
where $\mathcal{V}_{q} = \left\{ \mathbf{v}_{q_1}, \dots, \mathbf{v}_{q_2}, \dots, \mathbf{v}_{q_{c_q}} \right\}$ represents the embeddings of the nodes in $e_q$, and $\mathbf{e}_q \in \mathbb{R}^{h}$ is the embedding vector for the hyperedge $e_q$. Common node aggregation functions include:  \textbf{Sum} (summing the embedding vectors of the nodes along each dimension), \textbf{Mean} (averaging the embedding vectors of the nodes along each dimension), and \textbf{MaxMin} (computing the difference between the maximum and minimum values of the embedding vectors along each dimension) \cite{yadati2020nhp}.

The core task of the classifier is to use the embedding vector of a candidate hyperedge to predict whether it belongs to the target set $\mathcal{E}'$. Usually, The classifier can be defined as:
\begin{equation}
    p_q = \operatorname{Classifier}(\mathbf{e}_q),
\end{equation}
where $p_q$ is the likelihood that the candidate hyperedge belongs to $\mathcal{E}'$. In practice, the classifier is typically implemented using a feedforward neural network \cite{hornik1989multilayer}.

\section{Methods}
This paper proposes a negative sampling method named \textbf{HNS} (hyperedge negative sampling). The core idea of HNS is to inject positive sample information into negative samples in the high-dimensional embedding space, thereby generating more challenging negative samples. Specifically, the method consists of three main steps: Firstly, hypergraph neural networks are utilized to map hyperedges into dense embedding vectors, capturing high-order topological structures and node attribute information. Secondly, an embedding perturbation strategy is employed, which injects the embedding features of positive samples into the embedding vectors of simple negative samples, thus creating more difficult-to-discriminate negative samples in the embedding space. Finally, the generated negative samples are used to train the hyperedge prediction model, thereby improving the model's prediction accuracy.

\subsection{Hyperedge Representation}
Hyperedge representation aims to map hyperedges into a high-dimensional embedding space, where the similarity between nodes is measured by the distance between their vectors: Nodes that are similar in terms of attributes or structure are mapped close to each other, while dissimilar nodes are mapped far apart. This embedding approach transforms hyperedges from a complex and often discrete non-Euclidean space into a continuous Euclidean space, allowing the similarity between hyperedges to be quantified by directly computing distances. This provides a foundation for generating negative samples in subsequent steps. For example, if an HGNN encoder is chosen, the encoding process is defined as:
\begin{equation}
\mathbf{V} = \operatorname{HGNN}(\mathbf{H}, \mathbf{X}; \mathbf{\Theta}),
\end{equation}
where $\Theta$ represents the learnable parameters of HGNN. It is worth emphasizing that HNS is plug-and-play and does not impose specific requirements on the hypergraph encoder. Therefore, any hypergraph neural network capable of generating node embeddings can be combined with HNS.

Similar to ordinary graph neural networks, the fundamental idea of hypergraph neural networks is also to iteratively update embedding vectors by combining information from a node itself and its neighboring nodes. The mathematical framework is as follows: Initially, let $\mathbf{V}^{0} = \mathbf{X}$. For the $\kappa$-th layer, where $\kappa = 1, 2, \dots, K$, we have:
\begin{align}
	\mathbf{a}^{\kappa}_{v} &= \operatorname{AGGERATOR}^{\kappa}\left\{ \mathbf{V}^{\kappa-1}_{u}: u \in N(v) \right\}, \\
	\mathbf{V}^{\kappa}_{v} &= \operatorname{COMBINE}^{\kappa} \left\{ \mathbf{V}^{\kappa-1}_v, \mathbf{a}^{\kappa}_{v} \right\},
\end{align}
where the function $\operatorname{AGGERATE}$ aggregates information from the neighboring nodes of each node, and the function $\operatorname{COMBINE}$ updates the current node's embedding by combining the aggregated information from neighboring nodes with the current node's embedding. $N(v)$ is the set of neighboring nodes of node $v$. The final node embeddings $\mathbf{V}^{K}$ from the last layer can be considered the ultimate output of node embeddings.

After generating node embeddings, we can construct the representation of any candidate hyperedge through aggregation. Given a candidate hyperedge $e_q = \{ v_{q_1}, v_{q_2}, \dots, v_{q_{c_q}} \}$, we can aggregate the embeddings of these nodes to generate the embedding vector for hyperedge $e_q$. Previous study has shown that the \textit{MaxMin} function is effective for this purpose \cite{yadati2020nhp}, so we adopt this function (note that HNS or other negative sampling methods do not impose specific requirements on the choice of aggregation functions). The process is as follows:
\begin{align}
	\mathbf{\text{Max}} &= \left( \max_{i=q_1, q_2, \dots, q_{c_q}} \mathbf{v}_i^{(1)}, \max_{i=q_1, q_2, \dots, q_{c_q}}\mathbf{v}_i^{(2)}, \dots, \max_{i=q_1, q_2, \dots, q_{c_q}}\mathbf{v}_i^{(h)} \right), \\
	\mathbf{\text{Min}} &= \left( \min_{i=q_1, q_2, \dots, q_{c_q}} \mathbf{v}_i^{(1)}, \min_{i=q_1, q_2, \dots, q_{c_q}}\mathbf{v}_i^{(2)}, \dots, \min_{i=q_1, q_2, \dots, q_{c_q}}\mathbf{v}_i^{(h)} \right), \\
	\mathbf{\text{MaxMin}} &= \mathbf{\text{Max}} - \mathbf{\text{Min}}.
\end{align}
The embedding vector of the candidate hyperedge is then given by $\mathbf{e}_q=\operatorname{MaxMin}$.

\subsection{Negative Sample Generation}

\begin{figure}[ht]
    \centering
    \includegraphics[width=0.7\linewidth]{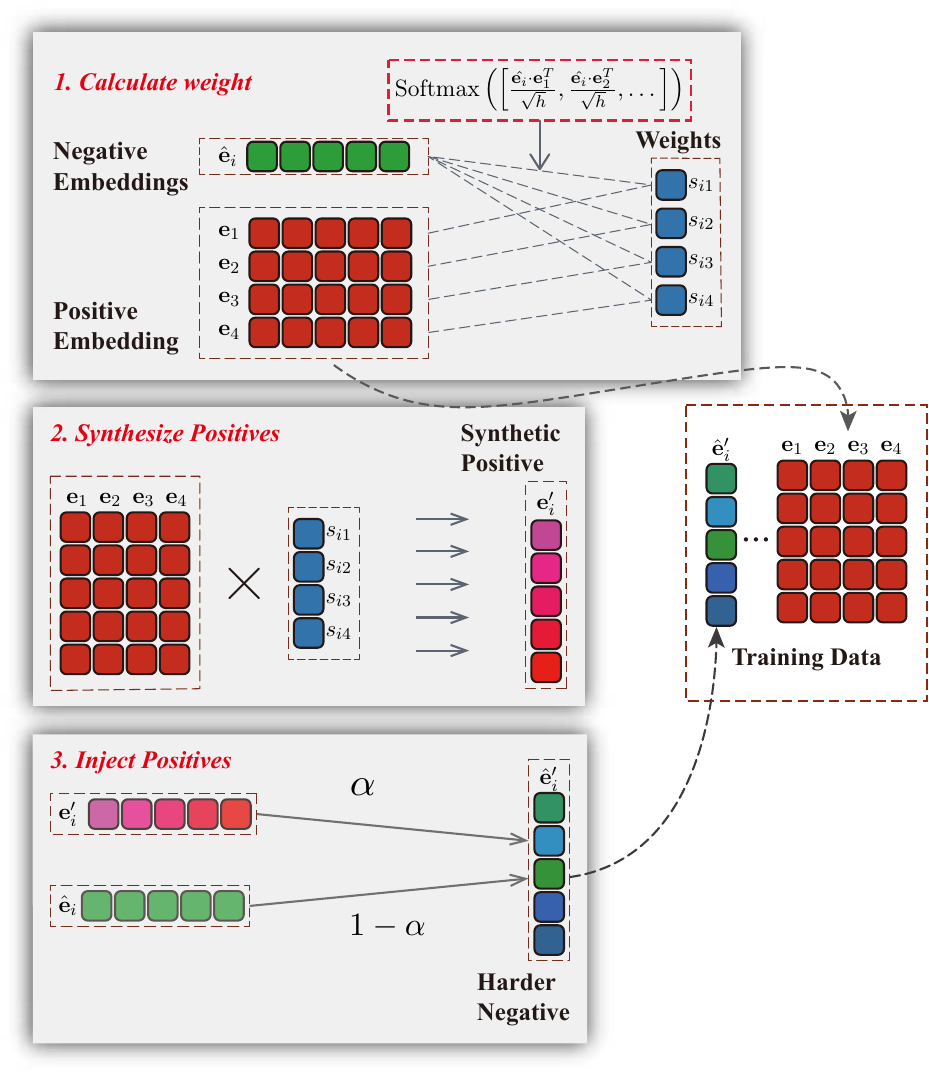}
    \caption{\textbf{Illustration of how to generate negative samples by HNS}. This framework operates through three sequential stages: (1) Weight Calculation: Compute the normalized weights between positive and negative sample embeddings, which will be further utilized to determine the contribution of each positive sample to the synthesis process. (2) Positive Sample Synthesis: Generating synthetic positive samples by weighted aggregation of multiple positive samples; and (3) Negative Sample Augmentation: Inject the synthetic positive samples into randomly sampled negative samples with a coefficient $\alpha$ to control the proportion of positive sample information.}
    \label{fig:inject}
\end{figure}

Compared to traditional heuristic hyperedge prediction algorithms, learning-based hyperedge prediction algorithms introduce a mapping function that embeds discrete nodes into a continuous vector space, allows the assessment of node similarity through vector distances. The core of this process lies in constructing an efficient mapping function that ensures similar nodes in the original network are close to each other in the vector space, while dissimilar nodes are far apart. To achieve this goal, the model must use both similar and dissimilar information during training. In hyperedge prediction tasks, it is typically assumed that nodes that have already formed or can form hyperedges are similar, while nodes that cannot form hyperedges are dissimilar. However, in a hypergraph, we can only observe the existing set of hyperedges $\mathcal{E}$, and we do not know which nodes cannot form hyperedges, making it impossible to obtain the necessary dissimilar information for building an efficient mapping function. Fortunately, real-world hypergraphs are extremely sparse, with the number of existent but unobserved hyperedges being far fewer than the possible hyperedges. Therefore, random negative sampling over all possible samples ensures a high probability that sampled negative samples are true negatives (\textit{i.e.}, they indeed cannot form hyperedges). However, as shown in Figure \ref{fig:embedding_visual}, randomly sampled negative samples are mostly easily distinguishable simple negatives, providing limited value for model training. Although some heuristic negative sampling methods exist, such as generating difficult-to-distinguish negative samples by merging multiple hyperedges or replacing one node within a hyperedge, these methods may generate false negatives (\textit{i.e.}, negative samples that actually constitute or should constitute hyperedges but have not been observed yet). Additionally, due to the different mechanisms of hyperedge formation across various scenarios, these heuristic methods lack generalizability.

To address the trade-off between authenticity and difficulty in negative sampling, the HNS method enhances the difficulty of negative samples by injecting the representations of positive samples into those of simple negative samples. As show in Figure \ref{fig:inject}, the core of the HNS method lies in performing positive sample injection on hyperedge representations, involving three key steps: Positive sample selection, positive sample synthesis, and hard negative sample generation.

Before injecting positive samples into negative samples, it is crucial to address the matching problem between positive and negative samples. Specifically, for a given simple negative sample, we need to determine which positive samples are suitable for injection. In the hyperedges embedding space, the positions of different positive samples relative to a particular negative sample determine their similarity, which in turn affects the difficulty of the generated negative samples. HNS employs a weighted selection strategy that considers influences of all positive samples for each negative by assigning different weights based on their distances (\textit{i.e.}, similarities to the target negative samples). Specifically, positive samples closer to the negative sample are assigned higher weights. This strategy is designed based on two considerations: Firstly, positive samples with shorter distances to the negative sample are typically closer to the classification decision boundary. The hyperedge prediction classifier aims to learn the boundary between positive and negative samples, and those positive samples near the negative sample are often located around this decision boundary. These samples are more valuable during model training, hence they should be given higher weights during injection. Secondly, positive samples that are farther from the negative sample generally exhibit greater differences in feature space. If such distant positive samples are given high weights during injection, their information may dominate the training process, potentially overshadowing or even replacing the features of the original negative sample. This could lead to the generation of false negative samples, misleading the classifier and degrading its discriminability.

To quantify the weights of positive samples relative to negative samples, HNS employs Scaled Dot-Product Attention \cite{vaswani2017attention}. Specifically, the weight $s_{ij}$ of a positive sample $\mathbf{e}_j$ for a given negative sample $\hat{\mathbf{e}}_i$ is calculated using the following formula:
\begin{equation}
	s_{ij} = \operatorname{softmax} \left(\frac{\hat{\mathbf{e}}_{i} \cdot \mathbf{e}_{j}^T}{\sqrt{h}} \right) 
= \frac{ \exp \left( \left( \hat{\mathbf{e}}_{i} \cdot \mathbf{e}_{j}^{T} \right) / \sqrt{h} \right) }{ \sum_{l=1}^{n} \exp \left( \left( \hat{\mathbf{e}}_{i} \cdot \mathbf{e}_{l}^{T} \right) / \sqrt{h} \right) }.
\end{equation}
Here, $1/\sqrt{h}$ is used to prevent gradient explosion or vanishing gradients, ensuring the stability during training. The $\operatorname{softmax}$ function \cite{bridle1990} normalizes the scores. Notably, the calculation of $s_{ij}$ is synchronized with the model's training process, meaning that $s_{ij}$ will be dynamically adjusted in the optimizing process of the model.

After computing the weights $s_{ij}$ of different positive samples $\mathbf{e}_{j}$ relative to a negative sample $\hat{\mathbf{e}}_{i}$, HNS synthesizes a corresponding positive sample ${\mathbf{e}}'_{i}$ for $\hat{\mathbf{e}}_{i}$ using linear weighting, as:
\begin{equation}
	\mathbf{e}'_{i} = \sum_{j=1}^{N} s_{ij} \mathbf{e}_{j}.
\end{equation}
Next, HNS employs a convex combination to inject the synthesized positive sample $\mathbf{e}'_{i}$ into the simple negative sample $\hat{\mathbf{e}_i}$, generating a hard negative sample $\hat{\mathbf{e}}'_{i}$:
\begin{equation}
	\hat{\mathbf{e}}'_{i} = (1 - \alpha) \hat{\mathbf{e}}_{i} + \alpha \mathbf{e}'_{i},
\end{equation}
where $\alpha \in (0, 1)$ controls the proportion of positive sample information in the synthesized hard negative sample. By adjusting the injection strength $\alpha$, one can precisely control the difficulty of the generated negative sample: The larger of $\alpha$, the harder it is to distinguish the generated negative sample. Specifically, $\alpha=0$ corresponds to no injection, and then the simple negative sample is used directly as the final negative sample. In this case, the HNS method degenerates into a random sampling method. The optimal value of $\alpha$ depends on different hypergraph datasets and hyperedge prediction algorithms, which can be determined experimentally.

\subsection{Training and Prediction}
The hard negative samples $\hat{\mathbf{e}}'_{i}$ generated by HNS, along with the original positive samples $\mathbf{e}_j$, form the training set $\mathcal{E}^{T}$, which is then fed into a multilayer perceptron (MLP)-based classifier \cite{bishop1995} for end-to-end training. This process jointly optimizes the parameters of both the encoder and the classifier. Specifically, for a candidate hyperedge $e_q$ (with its embedding representation denoted as $\mathbf{e}_q$), the model predicts its existence probability as:
\begin{equation}
	\hat{y}_{q} = \sigma \left( \mathbf{W}^{(2)} \cdot \text{ReLU} \left( \mathbf{W}^{(1)} \mathbf{e}_q + \mathbf{b}^{(1)} \right) + \mathbf{b}^{(2)} \right),
\end{equation}
where $\mathbf{W}^{(1)} \in \mathbb{R}^{h\times d}$ and $\mathbf{W}^{(2)} \in \mathbb{R}^{1\times h}$ are learnable weight matrices; $\mathbf{b}^{(1)}$ and $\mathbf{b}^{(2)}$ are bias terms; $\text{ReLU}(\cdot)$ is the activation function that introduces non-linearity to the model, and $\sigma(\cdot)$ is the Sigmoid function, which maps the logit values to probability values in the range $[0,1]$.

To enhance the model's robustness, we introduce an $L_2$ term into the loss function to prevent overfitting. The optimization objective of the model is defined as a weighted binary cross-entropy loss:
\begin{equation}
	\mathcal{L}_{\text{pred}} = -\frac{1}{|\mathcal{E}^{T}|} \sum_{q \in \mathcal{E}'} \left[  y_q \log \hat{y}_q + (1-y_q) \log(1-\hat{y}_q) \right] + \lambda \|\Theta\|_2^2,
\end{equation}
where $y_{q} \in \{0, 1\}$ is the true label of hyperedge $e_q$, $\hat{y}_{q}$ is the predicted probability for hyperedge $e_q$, $\lambda$ is the regularization strength, and $\Theta$ represents all learnable parameters. This loss function aims to maximize the prediction scores for positive samples and minimize the predicted scores for negative samples, thereby improving the model's discriminability. During the training process, the backpropagation algorithm \cite{rojas1996} is used to continuously adjust parameters of both the encoder and the classifier to minimize the loss function $\mathcal{L}_{\text{pred}}$.

\subsection{Algorithm Complexity}
When generating more challenging negative samples, the HNS method does not introduce additional model parameters, thus maintaining the same space complexity as the original model. Although some intermediate results are generated during the computation, they only temporarily occupy memory during calculation and are not stored long-term. Compared to the number of parameters in the model itself, the impact of these intermediate results on space complexity is negligible.

In terms of time complexity, the primary additional overhead of the HNS comes from the computation of weights $s_{ij}$ between positive and negative sample pairs. Specifically, for each generated simple negative sample, the model needs to compute its similarity with all positive samples. The time complexity of this process is $O(m_1 \times m_2 \times h)$, where $m_1$ represents the number of positive samples, $m_2$ represents the number of simple negative samples, and $h$ is the dimensionality of the hyperedge embedding representation. Since similarity calculations are required for every pair of positive and negative samples, this computation can become quite time-consuming on large hypergraph datasets. To reduce computational load and avoid the model getting stuck in local minima, neural networks typically employ batch training. Specifically, in each iteration, the model processes only a small batch of samples rather than the entire dataset \cite{krizhevsky2017imagenet}. Therefore, in the time complexity expression, $m_1$ and $m_2$ actually refer to the numbers of positive and negative samples within a batch, which are much smaller than the total number of samples in the entire dataset. This batch processing approach allows the HNS to maintain high performance even on large-scale datasets. Additionally, similarity computations can be implemented using matrix operations and can take full advantage of GPU acceleration, further reducing computational cost. This ensures that the HNS remains highly efficient and scalable in practical applications.

\section{Experiments}
To evaluate the effectiveness and robustness of HNS in hyperedge prediction tasks, we designed a series of experiments. Firstly, we compared the prediction accuracy of HNS with that of random negative sampling and state-of-the-art heuristic negative sampling methods using multiple evaluation metrics. Secondly, we conducted robustness experiments by analyzing the impact of different positive-negative sample matching strategies on model performance, and the effect of the injection strength $\alpha$ on model performance.

\subsection{Data Description}

\begin{table}[!ht]
    \centering
	\scriptsize
	\setlength{\tabcolsep}{7pt}
	\begin{tabular}{lrrrrrr}
		\toprule
		Dataset & $m$ & $n$ & $\left< k \right>$ & $\left< c \right>$ & $c_{max}$ & $d$ \\
		  \midrule
        Citeseer \cite{2008Cora}   & 1458 & 1079 & 2.3683 & 3.2002 & 26 & 3703  \\
        Cora \cite{2008Cora}        &  1434 &  1579 & 3.3375 & 3.0310 & 5 & 1433 \\
		Pubmed \cite{2012Pubmed}       &  3840   & 7963   & 9.0180 & 4.3487 & 362 & 500     \\
        Email-Enron \cite{Benson-2018-simplicial}  & 148 & 1544 & 33.0140 & 3.0576 & 37 &  148  \\
		NDC-Class \cite{Benson-2018-simplicial}   & 1149 & 1049 & 5.6292 & 6.1658 & 39 & 1149  \\
		Human-Disease \cite{goh2007human} & 330 & 516 & 1.8934 & 2.9606 & 11 & 22 \\
		Plant-Pollinator \cite{bhl43820} & 423 & 240 & 8.8958 & 5.0473 & 33 & 78 \\
		\bottomrule
	\end{tabular}
	\caption{\textbf{Statistical information of the datasets used in experiments}. There lists the following statistics for each dataset: The number of nodes $m$, the number of hyperedges $n$, the average degree of node $\left< k \right>$, the average size of hyperedge $\left< c \right>$, the maximum size of hyperedge $c_{max}$, and the dimensionality of node features $d$.}
\label{tab:dataset_info}
\end{table}

In the experiment, we select seven real hypergraphs from three different domains: Literature citation, social communication, and biomedicine. Citeseer \cite{2008Cora}, Cora \cite{2008Cora} and Pubmed \cite{2012Pubmed} were constructed based on citation co-occurrence relationships, where nodes represent academic papers and hyperedges represent sets of papers that are cited together in one paper. For node features, \text{Citeseer} and \text{Cora} use binary bag-of-words models to construct $3,703$-dimensional and $1,433$-dimensional binary feature vectors, respectively. Each dimension takes a value of either $0$ or $1$, indicating the presence or absence of specific words in the vocabulary. \text{Pubmed} uses TF-IDF \cite{salton1988} weighting to generate $500$-dimensional feature vectors, where each dimension is a non-negative real number. Email-Enron \cite{Benson-2018-simplicial} was built from internal email communication records of Enron Corporation, where hyperedges are defined as communication groups consisting of a single sender and all recipients. Since this dataset lacks inherent node features, we assigned each node a $148$-dimensional one-hot encoded vector, with each dimension uniquely corresponding to an employee's email address. A node's feature vector has a value of $1$ in its corresponding dimension and $0$ in the other $147$ dimensions. NDC-Class \cite{Benson-2018-simplicial} describes a drug classification system, where the hyperedges correspond to specific drug products, and nodes represent drug category labels. Node features are based on hierarchical one-hot encoding, resulting in $1,149$-dimensional binary vectors that directly indicate the drug category identity. Human-disease \cite{goh2007human} originates from a disease-gene association network, where nodes represent diseases and hyperedges connect sets of diseases sharing the same pathogenic genes. Node features include $22$-dimensional one-hot encoded vectors, with each dimension corresponding to a disease type (\textit{e.g.}, skeletal system diseases, cardiovascular diseases, and so on.). Plant-Pollinator \cite{bhl43820} records ecological interactions between plants and pollinators, where nodes represent plant species and hyperedges represent sets of plants associated with a particular pollinator. Node features use $78$-dimensional one-hot encoded vectors to indicate the biological classification of the plants. For data splitting strategies: Citeseer, Cora, Pubmed, Email-Enron, NDC-Class, and Plant-Pollinator adopt a split ratio of $60\%$ training set, $20\%$ validation set, and  $20\%$ test set. Due to the smaller sample size of the Human-disease dataset, a $6:2:2$ split would result in a fragmented training set. Therefore, we adjusted the split ratio to $8:1:1$ for this dataset, ensuring statistical efficacy in validation/test sets while maintaining the integrity of the hypergraph structure. Basic statistics of each hypergraph are summarized in Table \ref{tab:dataset_info}.

\subsection{Evaluation Metrics}
Recent studies have shown that different evaluation metrics exhibit only moderate correlation in binary classification problems \cite{bi2024inconsistency}. This implies that different metrics can often provide varying rankings of algorithms, and thus to use a single metric for evaluation may lead to biased conclusions. Therefore, this paper employs four different metrics to compare and analyze the prediction accuracy of various sampling methods. By combining the true hyperedge status in the network with the predictions made by the algorithm, we can derive the values in the confusion matrix. Assume that the set of positive samples in the test set is denoted as $\mathcal{E}^{P}$, and the set of negative samples in the test set is denoted as $\mathcal{E}^{N}$. In $\mathcal{E}^{P}$, the number of hyperedges correctly predicted as positive is True Positive (TP), while the number of hyperedges incorrectly predicted as negative is False Negative (FN). Similarly, in $\mathcal{E}^{N}$, the number of hyperedges correctly predicted as negative is True Negative (TN), while the number of hyperedges incorrectly predicted as positive is False Positive (FP). In this study, we utilize two evaluation metrics based on the confusion matrix:

The Area under the Receiver Operating Characteristic (ROC) Curve (AUC, also known as AUROC) is composed of coordinate points $(FPR, TPR)$ at different thresholds, where $FPR=\frac{FP}{FP+TN}$ denotes the False Positive Rate (FPR), and $TPR=\frac{TP}{TP+FN}$ denotes the True Positive Rate (TPR) \cite{hanley1982}. AUC can also be interpreted as the probability that a randomly selected hyperedge from $\mathcal{E}^{P}$ has a higher score $\hat{y}_{q}$ than a randomly selected hyperedge from $\mathcal{E}^{N}$. This can be expressed as: 
\begin{equation}
AUC=\frac{t_{1}+0.5t_{2}}{t},
\end{equation}
where $t_{1}$ is the number of times in $t$ independent comparisons that a hyperedge from $\mathcal{E}^{P}$ has a higher score than a hyperedge from $\mathcal{E}^{N}$, and $t_{2}$ is the number of times the scores are equal. The value of AUC ranges between $0$ and $1$. If predictions are made by random guessing, the AUC value is $0.5$. In contrast, if the predictions are perfect, the AUC value is $1$.

The Area under the Precision-Recall Curve (AUPR) is composed of coordinate points $(\text{Recall}, \text{Precision})$ at different thresholds \cite{buckland1994relationship, davis2006}. Here, \textbf{Recall} is the proportion of all hyperedges in $\mathcal{E}^{P}$ that are correctly predicted as positive, and \textbf{Precision} is the proportion of predicted positive hyperedges that actually belong to $\mathcal{E}^{P}$. AUPR can be calculated using the following formula \cite{bi2024inconsistency}: 
\begin{equation}
AUPR =\frac{1}{2 \cdot |\mathcal{E}^{P}|} \left(\sum\limits_{i=1}^{|\mathcal{E}^{P}|} \frac{i}{r_i} + \sum\limits_{i=1}^{|\mathcal{E}^{P}|} \frac{i}{r_{i+1}-1} \right),
\end{equation}
where $r_{i}$ denotes the rank of the $i$-th positive sample in the test set.

Additionally, we considered two evaluation metrics from the perspective of ranking quality. Normalized Discounted Cumulative Gain (NDCG) is a commonly used metric in information retrieval to evaluate the performance of ranking algorithms. This metric assigns higher weights to higher-ranked positions, thus capturing both the accuracy of the predictions and the reasonableness of the ranking \cite{jarvelin2002}. Intuitively, if a negative sample is incorrectly classified as positive, the higher its rank (\textit{i.e.}, the higher its score), the more it penalizes the NDCG metric. NDCG can be calculated using the following formula \cite{bi2024inconsistency}:
\begin{equation}
NDCG=\frac{\sum_{i=1}^{|\mathcal{E}^{P}|}\frac{1}{\log_2(1+r_i)}}{\sum_{r=1}^{|\mathcal{E}^{P}|}\frac{1}{\log_2(1+r)}}.
\end{equation}

Finally, we use the Mean Reciprocal Rank (MRR) metric that calculates the mean of the reciprocal ranks of the positive samples in the predicted ranking \cite{burges2005}. It measures how well the model can efficiently locate effective hyperedges, defined as:
\begin{equation}
MRR = \frac{1}{|\mathcal{E}^{P}|} \sum_{i=1}^{|\mathcal{E}^{P}|} \frac{1}{r_{i}}.
\end{equation}
The value of MRR ranges between $0$ and $1$. A higher MRR indicates better performance, meaning that the algorithm is more effective at ranking relevant hyperedges at higher positions.

AUC is one of the most popular evaluation metrics for hyperedge prediction \cite{yadati2020nhp, hwang2022ahp}. Although its effectiveness in handling imbalanced learning problems has been questioned, in this study, after negative sampling, the prediction task becomes a balanced learning problem. Furthermore, recent studies have shown that AUC has high discriminability \cite{zhou2023discriminating, jiao2024comparing, wan2024quantifying}. Therefore, the main text primarily presents results based on AUC, while results from other metrics are provided in the \ref{AppendixB}.

\subsection{Benchmarks}

To demonstrate the effectiveness of the proposed HNS method in the negative sampling task, we compared HNS with several other negative sampling methods in our experiments. The comparison methods were primarily drawn from three heuristic strategies proposed in \cite{patil2020negative}, namely Sized Negative Sampling (SNS), Motif Negative Sampling (MNS), and Clique Negative Sampling (CNS).

\begin{figure}[ht]
    \centering
    \includegraphics[width=0.7\linewidth]{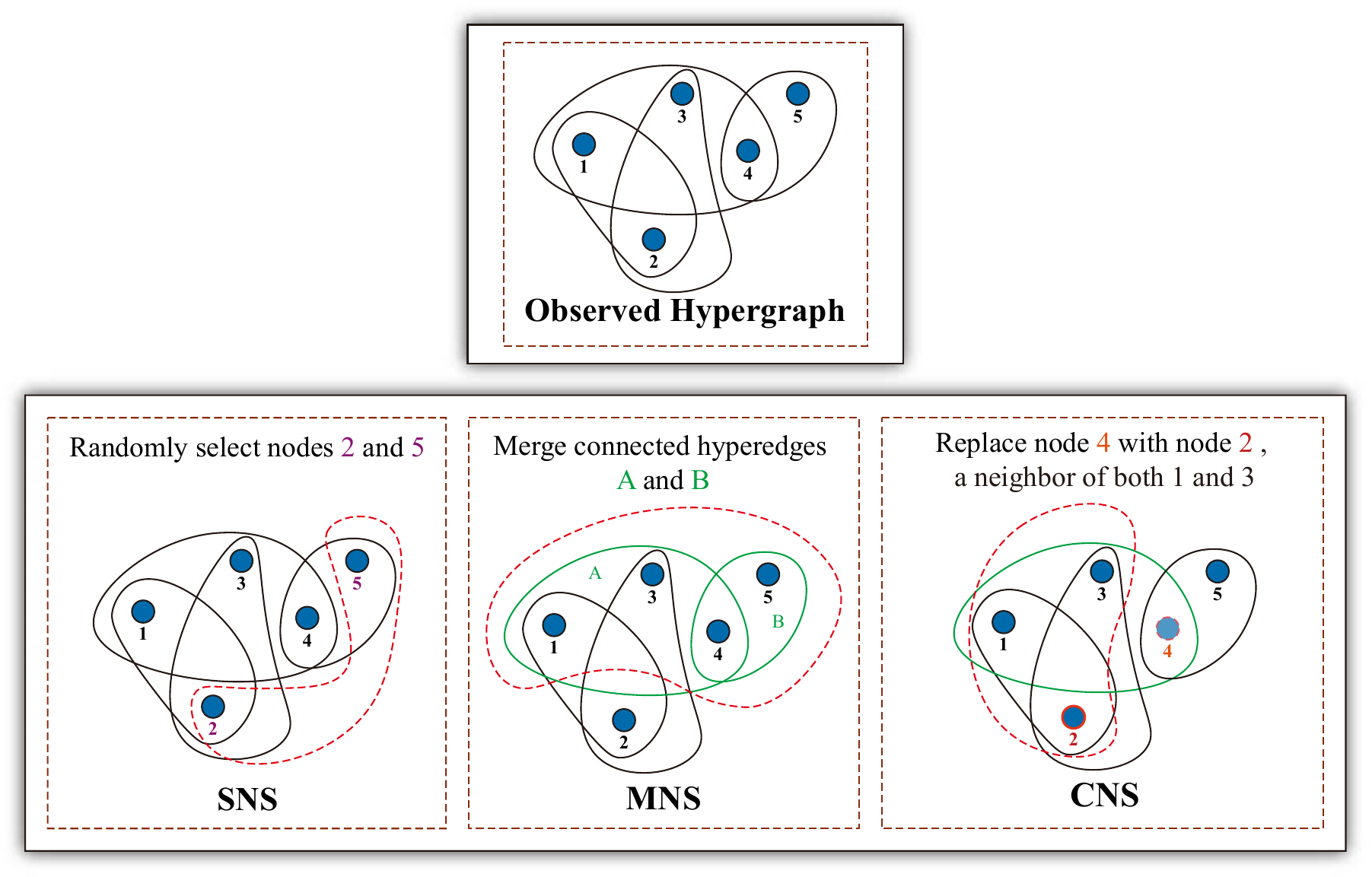}
    \caption{\textbf{Illustration of the three negative sampling methods for comparison}. }
    \label{fig:heuristic_hyperedge_negative_sampling}
\end{figure}

\begin{table}[ht]
    \centering
    \scriptsize
    \setlength{\tabcolsep}{4.5pt}
    \begin{tabularx}{\linewidth}{l*{7}{>{\centering\arraybackslash}X}}
    \toprule
    \Th{Model/Data} & \Th{CiteSeer} & \Th{Cora} & \Th{Pubmed} & \Th{NDC-Class} & \Th{Email-Enron} & \Th{Human-Disease} & \Th{Plant-Pollinator} \\
    \midrule
    HGNN(SNS) & 0.9333 & 0.8899 & \underline{0.9502} & \underline{0.9870} & 0.9755 & \underline{0.9636} & 0.9921 \\
    HGNN(MNS@V) & \underline{0.9558} & 0.8648 & 0.9374 & 0.9745 & 0.9748 & 0.9215 & \underline{0.9924} \\
    HGNN(CNS@V) & 0.9284 & 0.8567 & 0.9446 & 0.9745 & \underline{0.9766} & 0.9581 & 0.9920 \\
    HGNN(MNS@T) & 0.8434 & \underline{0.9221} & 0.7497  & 0.9777 & 0.9590 & 0.9625 & 0.9901 \\
    HGNN(CNS@T) & 0.5887 & 0.4330 & 0.3265 & 0.7595 & 0.8304 & 0.4582 & 0.7008 \\
    HGNN(HNS) & \textbf{0.9597} & \textbf{0.9537} & \textbf{0.9661}  & \textbf{0.9881} & \textbf{0.9806} & \textbf{0.9706} & \textbf{0.9928} \\
    \midrule
    HNHN(SNS) & \underline{0.9573} & \underline{0.9342} & \underline{0.9415} & \underline{0.9931} & \underline{0.9781} & 0.9627 & \textbf{0.9789} \\
    HNHN(MNS@V) & 0.9540 & 0.8977 & 0.9367 & 0.9843 & 0.9715 & 0.9450 & \underline{0.9774} \\
    HNHN(CNS@V) & 0.9505 & 0.8808 & 0.8460 & 0.9841 & 0.9706 & \underline{0.9647} & 0.8956 \\
    HNHN(MNS@T) & 0.8840 & 0.8762 & 0.8324 & 0.9198 & 0.9627 & 0.9619 & 0.9597 \\
    HNHN(CNS@T) & 0.7033 & 0.4837 & 0.2629 & 0.9150 & 0.8808 & 0.3399 & 0.8336 \\
    HNHN(HNS) & \textbf{0.9601} & \textbf{0.9451} & \textbf{0.9426} & \textbf{0.9940} & \textbf{0.9823} & \textbf{0.9657} & 0.9772 \\
    \midrule
    NHP(SNS) & 0.9589 & 0.9137 & \underline{0.9414} & \underline{0.9974} & 0.9780 & \underline{0.9570} & \underline{0.9746} \\
    NHP(MNS@V) & \underline{0.9628} & \underline{0.9291} & 0.9326 & 0.9952 & \underline{0.9781} & 0.9219 & \textbf{0.9757} \\
    NHP(CNS@V) & 0.9480 & 0.9124 & 0.9240 & 0.9887 & 0.9762 & 0.9273 & 0.7374 \\
    NHP(MNS@T) & 0.8681 & 0.9261 & 0.5252 & 0.9748 & 0.9680 & 0.9343 & 0.8866 \\
    NHP(CNS@T) & 0.0702 & 0.1558 & 0.2932 & 0.6595 & 0.7618 & 0.1475 & 0.5846 \\
    NHP(HNS) & \textbf{0.9638} & \textbf{0.9359} & \textbf{0.9415} & \textbf{0.9977} & \textbf{0.9814} & \textbf{0.9620} & 0.9741 \\
    \midrule
    HGNNP(SNS) & \underline{0.9719} & \underline{0.9172} & \textbf{0.9406} & \underline{0.9963} & \underline{0.9845} & \underline{0.9550} & \textbf{0.9797} \\
    HGNNP(MNS@V) & 0.9539 & 0.8886 & 0.9325 & 0.9911 & 0.9839 & 0.8825 & \underline{0.9785} \\
    HGNNP(CNS@V) & 0.9605 & 0.8977 & 0.9319 & 0.9867 & 0.9840 & 0.8790 & 0.8709 \\
    HGNNP(MNS@T) & 0.9127 & 0.8580 & 0.8433 & 0.9654 & 0.9666 & 0.9535 & 0.9638 \\
    HGNNP(CNS@T) & 0.7918 & 0.3815 & 0.2058 & 0.9630 & 0.8783 & 0.2195 & 0.8347 \\
    HGNNP(HNS) & \textbf{0.9828} & \textbf{0.9308} & \underline{0.9354} & \textbf{0.9965} & \textbf{0.9857} & \textbf{0.9574} & 0.9775 \\
    \bottomrule
    \end{tabularx}
    \caption{\textbf{Performance comparison of hypergraph negative sampling strategies across multiple datasets and model architectures, subject to AUC}. The best and second-best results are highlighted in \textbf{black bold} and \underline{underlined}, respectively.}
    \label{tab_auc_performance}
    \end{table}

The basic idea of the SNS method is to construct candidate negative samples by randomly selecting a predetermined number of nodes, ensuring that the generated hyperedges are not in the observed positive sample set. For example, as illustrated in Figure \ref{fig:heuristic_hyperedge_negative_sampling}, if the target hyperedge size is $2$, two nodes (\textit{e.g.}, nodes $2$ and $5$) are randomly chosen from the node set $\{1,2,3,4,5\}$, forming the hyperedge $\{2,5\}$. In practice, the distribution of hyperedge sizes in the negative samples is typically set to be consistent to the distribution of hyperedge sizes in the positive samples. This means that the size of each negative sample can be obtained by independently sampling from the distribution of hyperedge sizes in positive samples. The method is computationally simple and efficient, making it suitable for large-scale hypergraphs. Because the candidate sampling set covers the entire possible hyperedge set, given the sparsity characteristic of real networks, the sampled negative samples are likely to be true negatives. However, since this method completely ignores the internal structural information of the hypergraph, the generated negative samples may significantly differ in structure from the existent hyperedges, making the negative samples overly simplistic and less valuable for improving model performance.

We define two hyperedges as mutual neighbors if they share at least one common node. The MNS method begins by randomly selecting a positive hyperedge as the starting point, then it randomly selects other neighboring positive hyperedges. It continuously merges the nodes from these neighboring hyperedges into the current set until the generated hyperedge reaches the predetermined size. For example, in Figure \ref{fig:heuristic_hyperedge_negative_sampling}, we first randomly select the hyperedge $\{1, 3, 4\}$ from the positive samples as the starting point, and then randomly choose a neighboring hyperedge  $\{4, 5\}$ from the set of neighboring hyperedges. By merging these two hyperedges, we construct a negative sample $\{1,3,4,5\}$. The negative samples generated in this manner are structurally closer to positive hyperedges, making them more challenging for classification tasks. However, due to the similarity between this merging process and the mechanism in generating positive samples, there is a risk of introducing false-negative samples, which can affect the authenticity of negative samples.

The CNS method constructs negative samples by replacing one node in a positive hyperedge. Specifically, the method first randomly selects a positive sample hyperedge, then randomly chooses one of its nodes, and finally replaces it with a node randomly selected from the common neighbors of the remaining nodes. For example, as illustrated in Figure \ref{fig:heuristic_hyperedge_negative_sampling}, suppose we randomly select node $4$ from the hyperedge $\{1, 3, 4\}$ for replacement. The common neighbors of the remaining nodes $1$ and $3$ are nodes $2$ and $4$. Since node $4$ is to be replaced, only node $2$ can be chosen. Therefore, the resulted negative hyperedge is $\{1,2,3\}$. This strategy largely preserves the structural characteristics of the original hyperedge while introducing only minor perturbations, thus generating structurally similar negative samples that are highly challenging for classification tasks. However, this method also has some practical drawbacks: Firstly, similar to MNS, it may generate false negative samples. Secondly, when the hypergraph is sparse, the set of common neighbors of the remaining nodes may be empty, making it impossible to generate valid negative hyperedges. Last but not least, this method requires traversing the neighbor sets of all nodes in the hyperedge, leading to significant computational overheads on large datasets.

In our experiments, we used different negative sampling methods to generate both the training set and the validation set. We use the SNS method as the baseline, which means that, unless otherwise specified, SNS is the default negative sampling method. When an experiment is labeled as SNS, it indicates that both the training set and the validation set are generated using the SNS method. Our proposed HNS method is primarily applied during the generation of the training set, while for the validation set generation, we simply use SNS. We also applied the MNS and CNS methods to either the training set or the validation set generation, resulting in four different comparison methods: (1) MNS@T, using the MNS method for training set generation; (2) MNS@V, using the MNS method for validation set generation; (3) CNS@T, using the CNS method for training set generation; and (4) CNS@V, using the CNS method for validation set generation. Experiments were conducted on four different network architectures: HGNN \cite{feng2019hypergraph}, HNHN \cite{dong2020hnhn}, NHP \cite{yadati2020nhp}, and HGNNP \cite{gao2023hgnn+}.

Table \ref{tab_auc_performance} presents the experimental results using AUC as the evaluation metric. The results for other metrics are very similar and can be found in \ref{AppendixB}. From Table \ref{tab_auc_performance}, we can observe the following four phenomena: (1) In $28$ experiments, HNS achieved the best performance in $24$ experiments, demonstrating its significantly better overall performance compared to other negative sampling methods. This advantage stems from that fact that HNS can adaptively improve the quality of negative samples as the embeddings of positive and negative samples are continuously optimized during training. Furthermore, HNS can control the difficulty of the generated negative samples by adjusting the hyperparameter $\alpha$, tailoring them to match the learning requirements of different models and the characteristics of different hypergraphs. (2) MNS@T and MNS@V outperform CNS@T and CNS@V. This is likely due to the fact that the CNS method generates negative samples by replacing only one node in a positive hyperedge. Such generated negative samples might still resemble positive samples or conform to the mechanisms used to generate positive samples, leading to false negative samples. These false negatives can negatively impact the performance of the model. (3) MNS@T and CNS@T perform poorly in some experiments, with results sometimes even worse than random guessing. We believe this is because, during training, the model learns the distribution patterns of both positive and negative samples simultaneously. When using negative samples generated by MNS or CNS, the model implicitly fits the distribution characteristics inherent to these methods. Since the distribution patterns of negative samples generated by MNS and CNS may highly overlap with those of positive samples, this ultimately disrupts the model's ability to distinguish between positive and negative samples, sometimes even misclassifying the real positive samples as negatives. (4) Compared to other methods, HNS exhibited more stable performance in experiments. Even when it did not achieve optimal performance, its results remained relatively high. This stability underscores the robustness of the HNS method in different scenarios.

\subsection{Robustness Analysis}

\begin{figure}[!p]
	\centering
	\includegraphics[width=0.8\textwidth]{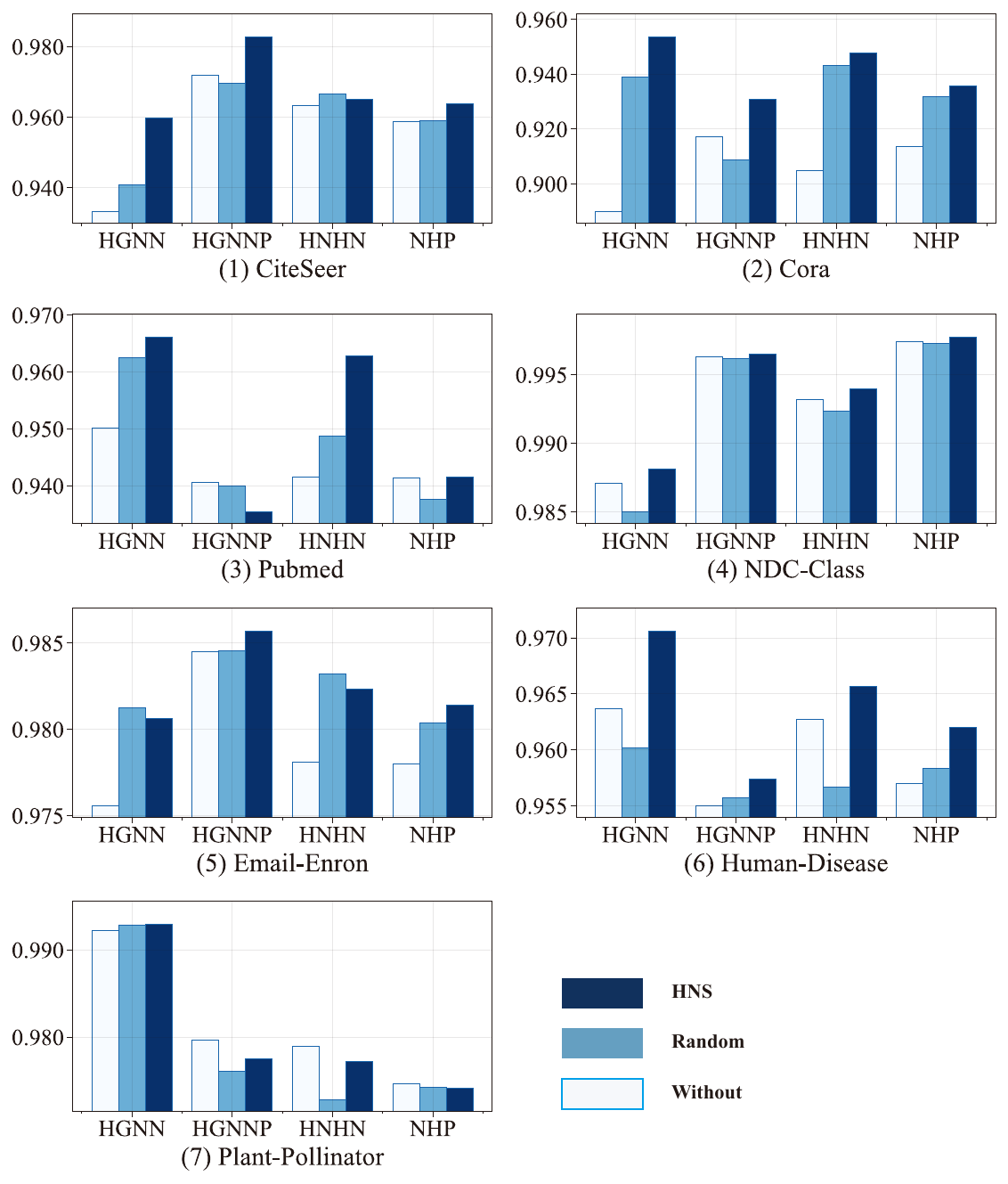}
	\caption{\textbf{Comparison of model performance under different positive sample selection strategies}. The X-axis represents different hypergraph neural networks, and the Y-axis shows the values of AUC. Subplots (1) to (7) represent the experimental results for the  seven real hypergraphs.}
	\label{fig:ablation}
\end{figure}

\begin{figure}[!p]
	\centering
	\includegraphics[width=0.5\textwidth]{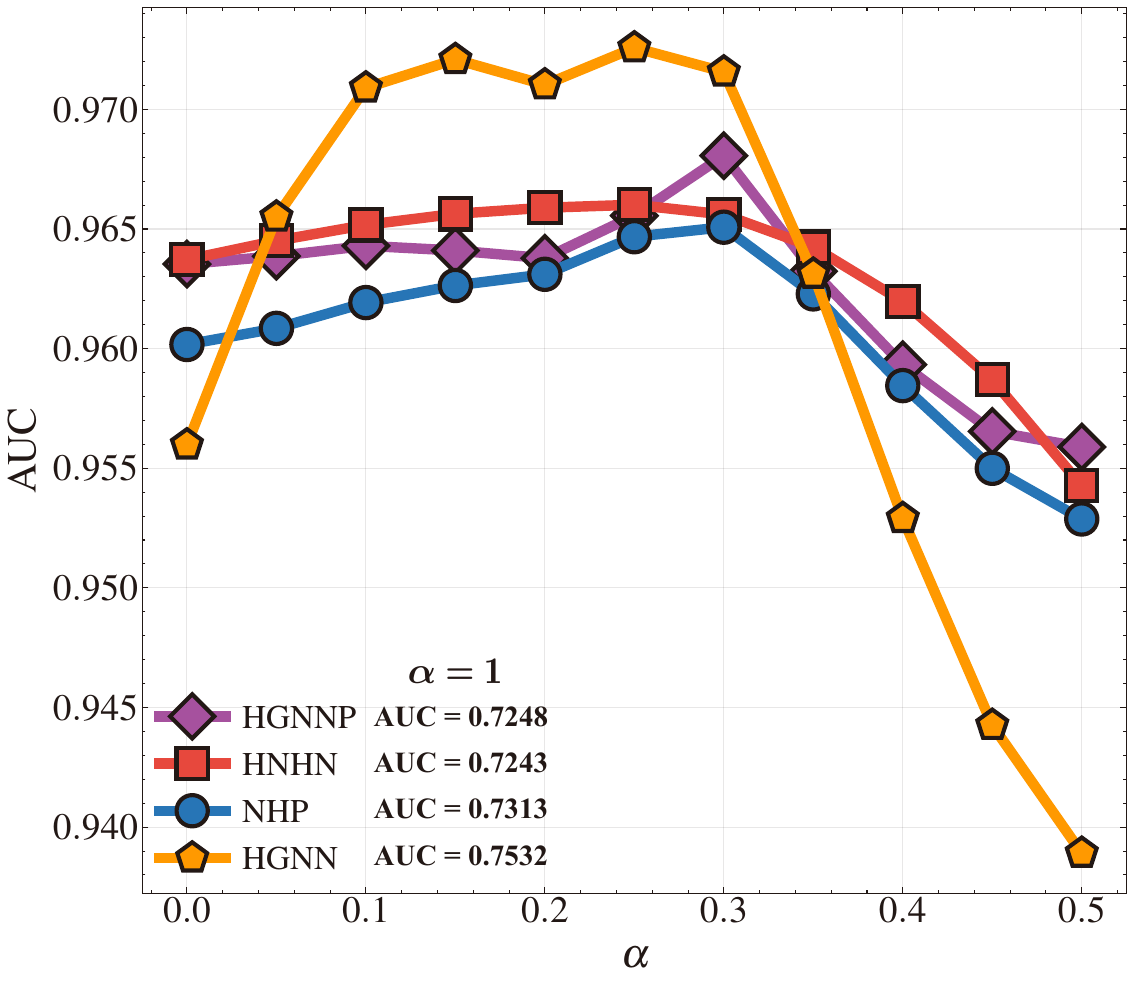}
	\caption{\textbf{The average AUC values with different $\alpha$}. Four colors represent four different methods. The figure also shows the AUC values for each method when $\alpha=1$.}
	\label{fig:by_method}
\end{figure}

\begin{figure}[!p]
	\centering
	\includegraphics[width=0.8\textwidth]{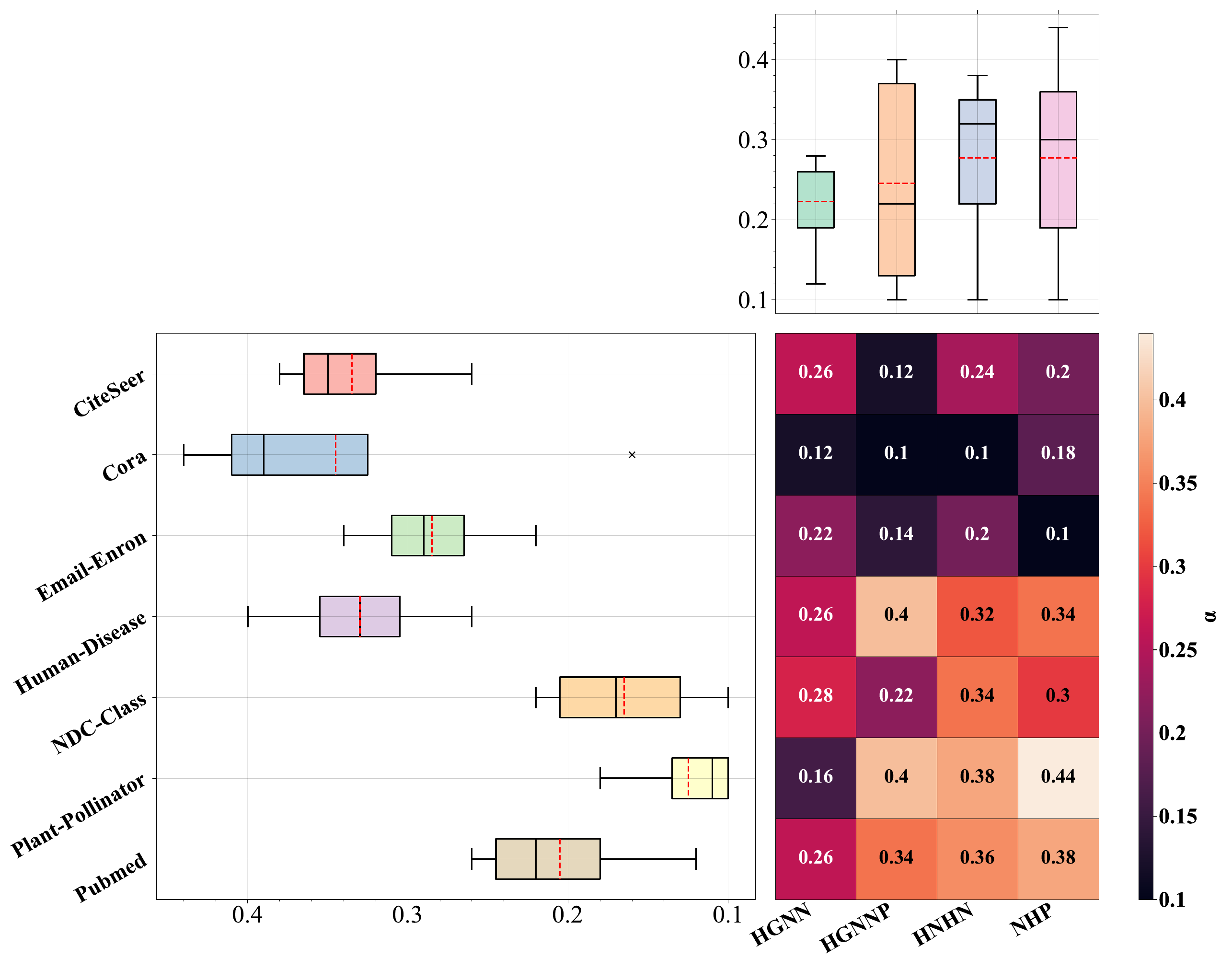}
	\caption{\textbf{Optimal hyperparameter $\alpha$ across different datasets and hyperedge prediction methods}. The central part of the figure is a heatmap where rows correspond to different datasets and columns correspond to different hyperedge prediction methods. The values in the heatmap indicate the optimal $\alpha$ values for each method on each dataset. Above and to the left of the heatmap are box plots showing the distribution of $\alpha$ values for different methods and datasets. In these box plots, solid lines represent the median, and red dashed lines represent the mean.}
	\label{fig:sensitivity}
\end{figure}

To evaluate the robustness of the HNS method, we explore it from two perspectives. Firstly, we compare the impact of different positive sample selection strategies on the final prediction results. This comparison aims to validate the rationality of using a similarity-based selection strategy in weighting positive samples in HNS. By examining how various selection strategies influence performance, we can better understand the effectiveness and reliability of the similarity-based approach. Secondly, We investigate the influence of the hyperparameter $\alpha$ on prediction performance. Specifically, we analyze how the optimal value of $\alpha$ varies across different hypergraph neural network models and datasets. This provides valuable insights and guidelines for selecting appropriate values of $\alpha$ under various conditions, ensuring optimal performance for different scenarios. By conducting these analyses, we aim to provide a comprehensive understanding of the robustness of HNS and offer practical guidance for its application in diverse settings.

To further investigate the impact of selecting positive hyperedges on the HNS, we introduced a variant of HNS that uses a random positive hyperedge selection strategy for comparison. Specifically, we randomly select one hyperedge from the set of positive hyperedges and inject it into the candidate negative hyperedge set, and then use grid search to select the weight coefficients that yield the best performance on the validation set. Additionally, we used a baseline method that does not employ any specific injection strategy as the baseline for comparison. These two comparative methods are labeled as “Random” and “Without”, respectively. The latter is equivalent to SNS in Table \ref{tab_auc_performance}. As shown in Figure \ref{fig:ablation}, the method using a similarity-based selection strategy for positive sample (labeled as HNS) achieved the best performance in $21$ out of $28$ tests. Moreover, both the method using a similarity-based selection strategy (HNS) and the one using a random selection strategy (Random) performed overall better than the method without any negative sample processing (SNS). This further validates the effectiveness of introducing hard negative samples for improving model performance.

The hyperparameter $\alpha$ represents the proportion of positive sample information in the synthesized negative samples. The optimal value of $\alpha$ to some extent reflects how challenging the negative samples need to be for a given dataset and a given algorithm. Figure \ref{fig:by_method} shows the average AUC values obtained from four different hypergraph neural network models as the parameter $\alpha$ varies. Here, $\alpha=0$ indicates that no negative samples are injected. When $\alpha$ approaches $1$, the high similarity between negative and positive samples leads to false negative effects, making it difficult for the model to correctly distinguish between positive and negative samples, which significantly degrades prediction performance. To avoid making the fluctuations in AUC at smaller $\alpha$ values difficult to observe due to very low AUC values when $\alpha=1$, we did not plot the range where $\alpha > 0.5$ (AUC with monotonically decrease from $0.5$ to $1$). Instead, we only show the AUC values at $\alpha=1$ in this figure. As shown in Figure \ref{fig:by_method}, the AUC changes with $\alpha$ in a unimodal manner; the predictive performance is poor at both extremes ($\alpha=0$ and $\alpha=1$). In cases where the specific network structure is unknown and $\alpha$ cannot be determined based on the network, for different hypergraph neural networks, the search range for the hyperparameter $\alpha$ can be set between $0.2$ and $0.4$ as suggested by Figure \ref{fig:by_method}. The optimal $\alpha$ value can then be determined through grid search \cite{bergstra2012} on the validation set. Figure \ref{fig:sensitivity} directly presents the optimal $\alpha$ values for given hypergraph neural network models and datasets. Compared to the sensitivity to the hypergraph neural network models used for representation learning, the optimal $\alpha$ value is more sensitive to the dataset. This highlights the value of introducing the hyperparameter $\alpha$, that is, by adjusting $\alpha$, one can modulate the difficulty of predicting negative samples to better fit different datasets.

\section{Conclusion and Discussion}
This paper proposes a negative sampling method named HNS, which is based on hyperedge embeddings. Unlike traditional negative sampling methods, HNS synthesizes hard-to-classify negative samples directly in the embedding space by injecting positive sample information into negative samples. Experimental results demonstrate that compared to existing hyperedge negative sampling methods, HNS can significantly enhance the performance of hyperedge prediction models. The core advantage of HNS lies in its flexibility and generality. As a plug-and-play method, HNS does not depend on specific model architectures or task scenarios, allowing it to be seamlessly integrable into most embedding-based hypergraph link prediction models. As long as the model can generate node embeddings, HNS can construct high-quality negative samples based on these embedding vectors. Furthermore, by synthesizing negative samples in the embedding space, HNS greatly expands the construction space for negative samples, theoretically enabling the generation of an infinite number of negative samples. Because it is not constrained by graph space, the HNS method or its variants can generate negative samples with specific requirements according to the needs of the problem. This provides richer and more diverse training samples for the model.

Although HNS demonstrates outstanding performance, it still has some limitations. Firstly, the effectiveness of HNS heavily relies on the quality of the embedding representations. If the embedding space fails to adequately capture the structural information, the generated negative samples may not effectively enhance model performance. This underscores the importance of high-quality embeddings in ensuring the success of HNS. Secondly, HNS introduces a hyperparameter $\alpha$ to regulate the difficulty of classifying negative samples. This parameter allows HNS to generate appropriately challenging negative samples based on the requirements of different tasks and datasets, thereby enhancing its flexibility and adaptability. However, this also increases the complexity of using HNS, as the parameter $\alpha$ needs to be tuned for specific tasks. The necessity to adjust $\alpha$ can add to the overall complexity of applying the method in various scenarios. In future research, we plan to explore methods for adaptively adjusting $\alpha$. By developing techniques that can automatically optimize this parameter, we aim to further improve the usability and robustness of this method. This will make HNS more accessible and effective in a wider range of applications.

In summary, HNS provides a novel and effective solution to sample negatives in hyperedge prediction. Unlike traditional negative sampling methods, the uniqueness of HNS lies in two key aspects: Generating negative samples through synthesis rather than simple sampling, and synthesizing these samples in the embedding space rather than in the original graph space. Given the plug-and-play characteristic of HNS, these innovative concepts are not limited to the specific problem of hyperedge prediction. We believe that both the specific methodologies and the perspective underlying HNS can be easily extended to other graph learning tasks, such as node classification, community detection, graph classification, and more. This versatility makes HNS a promising tool for a wide range of applications in graph-based machine learning.





\newpage
\appendix

\setcounter{table}{0}   
\setcounter{figure}{0}
\setcounter{section}{0}
\setcounter{equation}{0}
\renewcommand{\thetable}{A\arabic{table}}
\renewcommand{\thefigure}{A\arabic{figure}}
\renewcommand{\theequation}{A\arabic{equation}}

\renewcommand{\thesection}{Appendix \Alph{section}}  


\section{Visual Comparison of Different Samples}\label{AppendixA}
\begin{figure}[!ht]
    \centering
    \subfigure[CiteSeer(top 20 nodes)]{
        \includegraphics[width=0.40\textwidth]{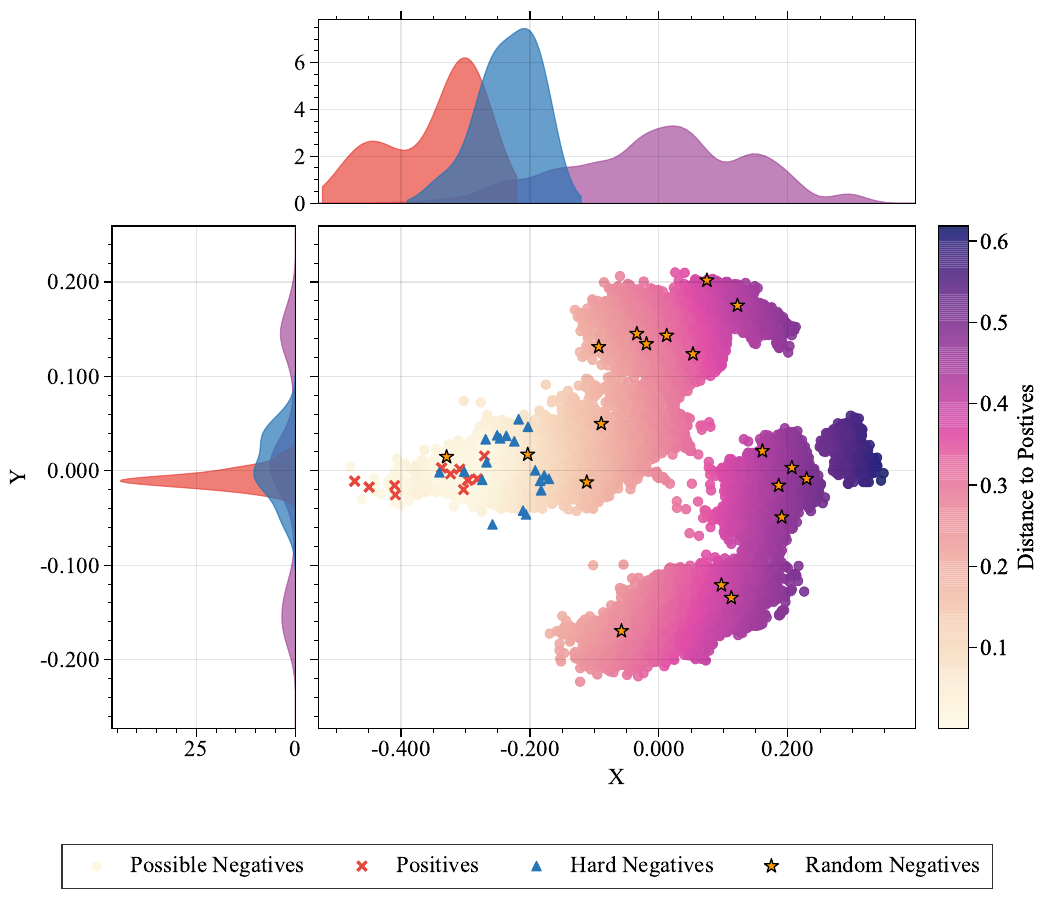}
    }
    \subfigure[Cora(top 27 nodes)]{
        \includegraphics[width=0.40\textwidth]{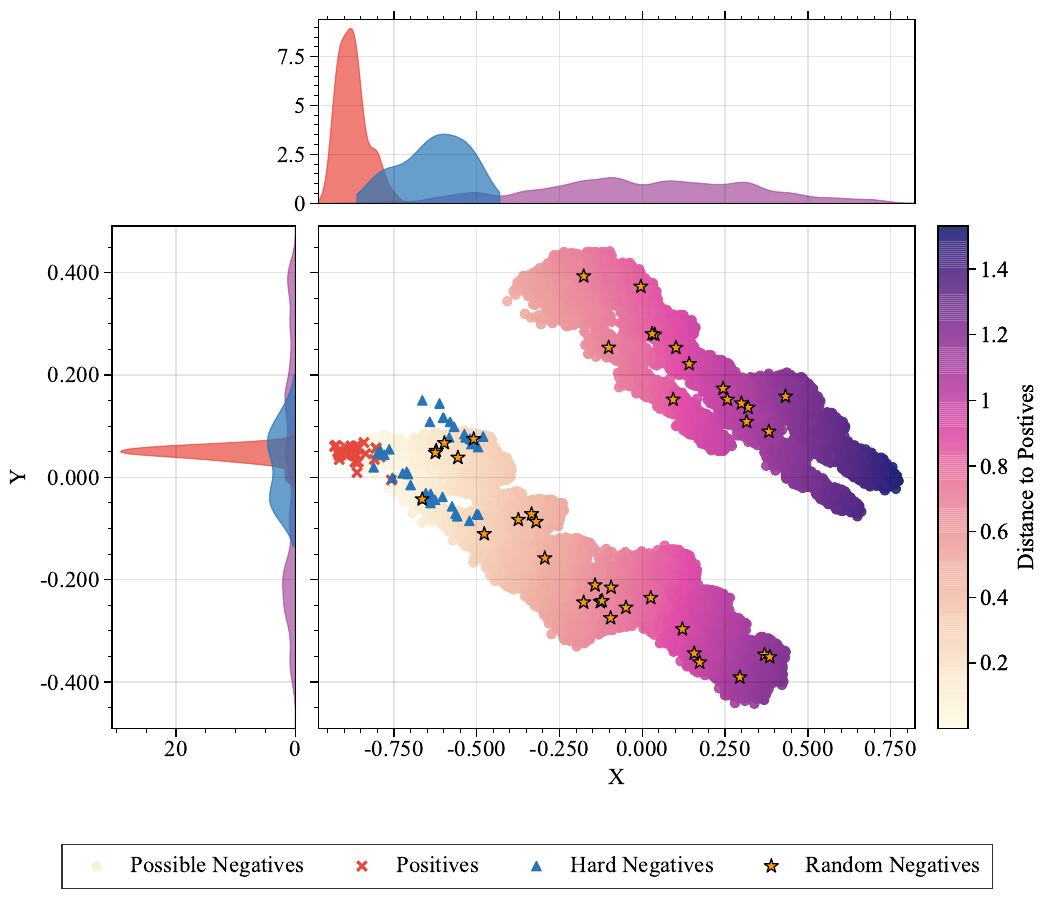}
    } \\
    \subfigure[Pubmed(top 17 nodes)]{
        \includegraphics[width=0.40\textwidth]{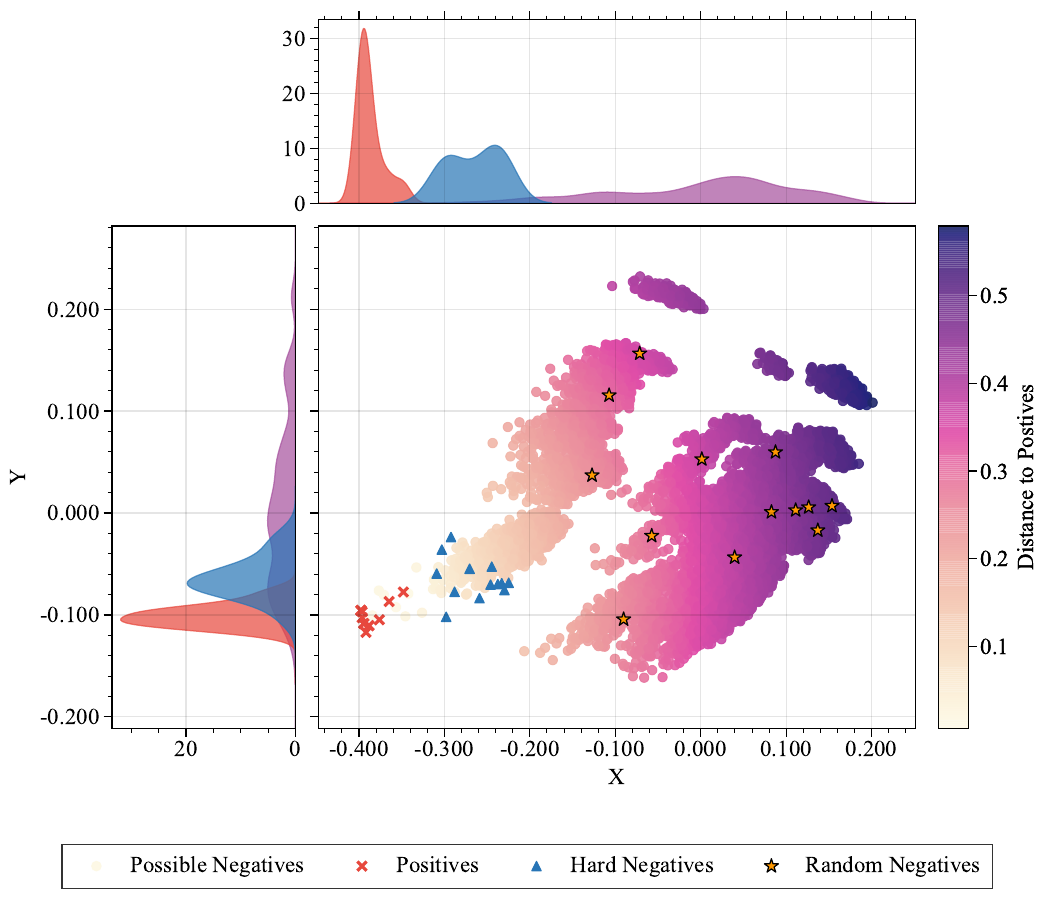}
    }
    \subfigure[NDC-Class(top 28 nodes)]{
        \includegraphics[width=0.40\textwidth]{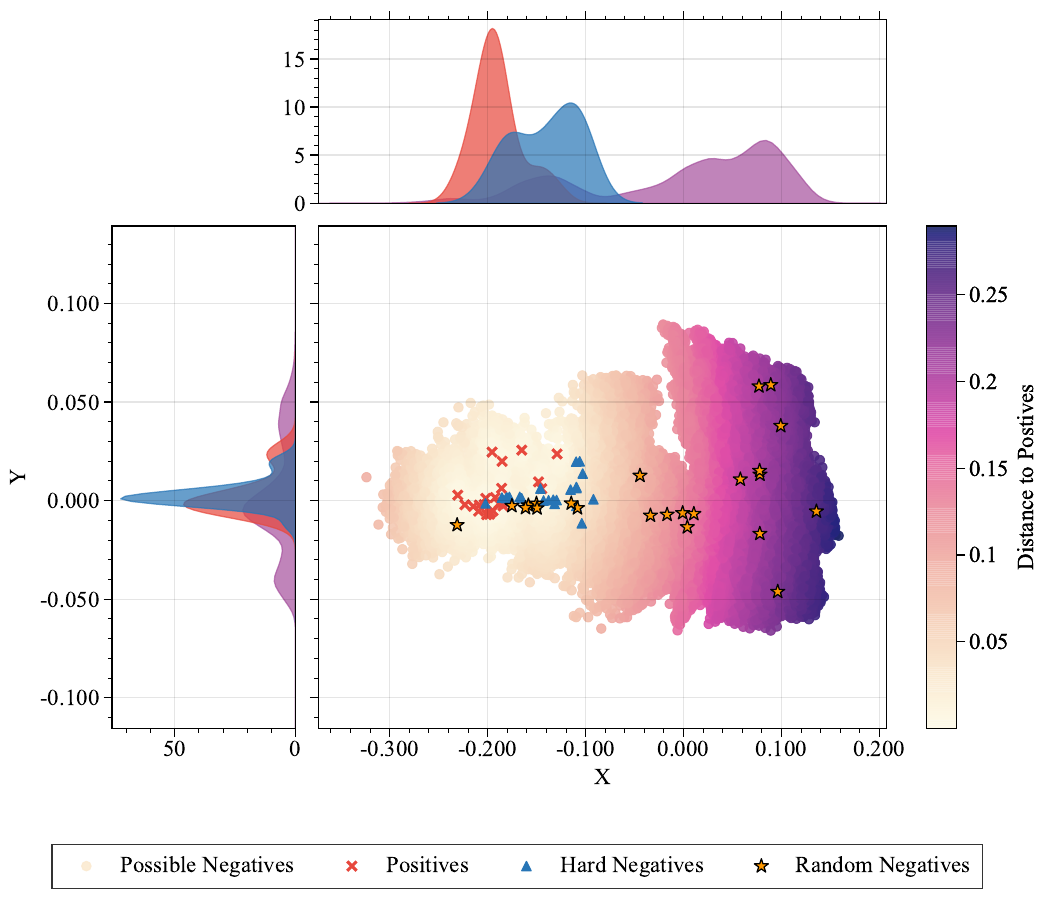}
    } \\
    \subfigure[HumanDisease(top 20 nodes)]{
        \includegraphics[width=0.40\textwidth]{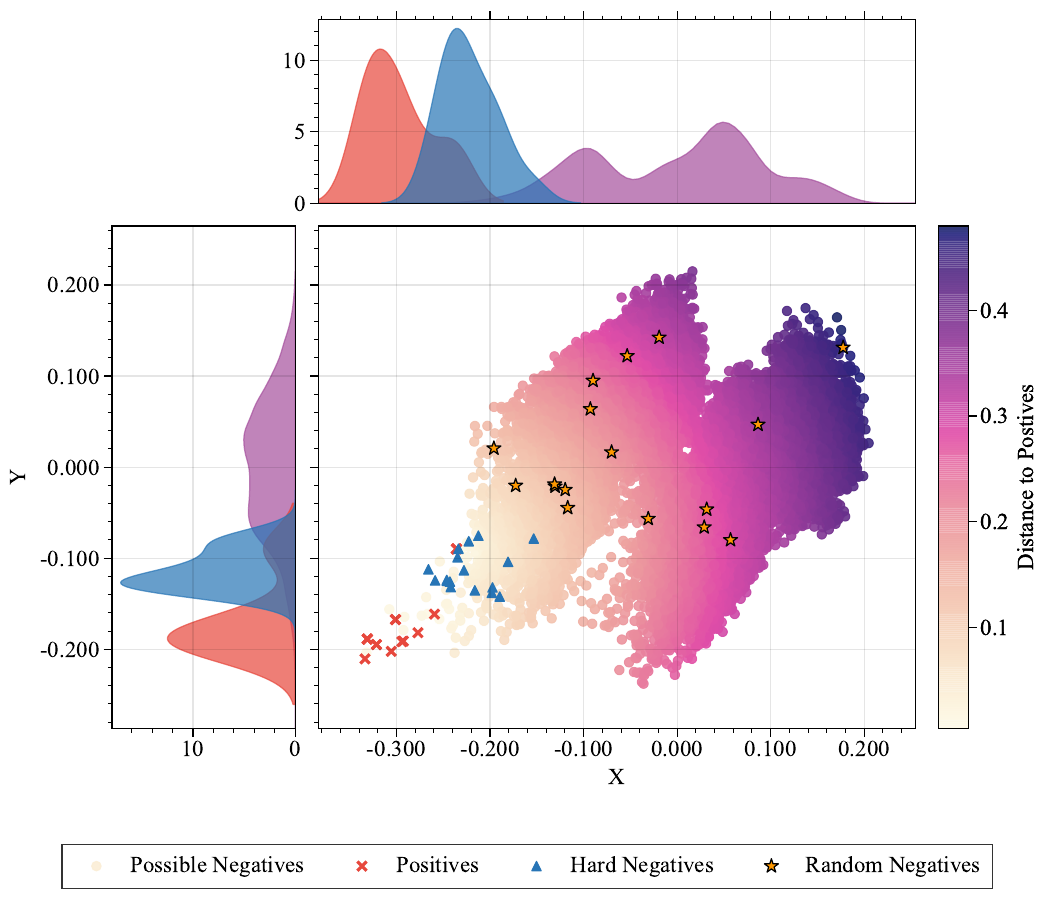}
    }
    \subfigure[SpeciesInteraction(top 26 nodes)]{
        \includegraphics[width=0.40\textwidth]{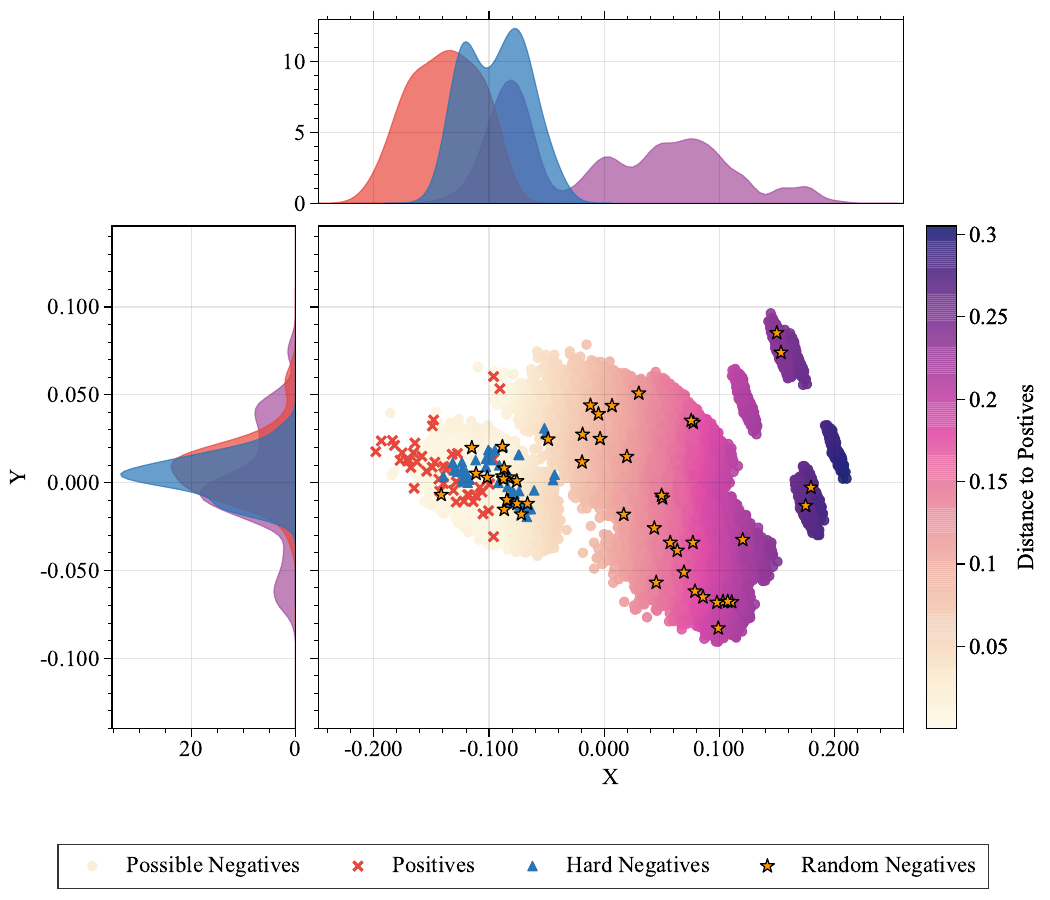}
    }
    \caption{\textbf{Visualization of hyperedge embeddings in reduced dimensionality}. Using self-supervised learning methods, we performed representation learning on hyperedges from high-frequency subgraphs across various datasets and projected the high-dimensional embeddings into a $2D$ space. The figure displays the distribution of four types of samples: (1) Possible Negatives: All potential negative samples with orders ranging from $3$ to $5$; (2) Positives: The set of positive samples; (3) Random Negatives: Negative samples randomly selected from all possible negatives and (4) Harder Negatives: negative samples obtained by the HNS method.}
    \label{fig:experiment_visualization}
\end{figure}

\section{Results for Alternative Evaluation Metrics}\label{AppendixB}

    \begin{table}[ht]
    \centering
    \scriptsize
    \setlength{\tabcolsep}{4.5pt}
    \begin{tabularx}{\linewidth}{l*{7}{>{\centering\arraybackslash}X}}
    \toprule
    \Th{Model} & \Th{CiteSeer} & \Th{Cora} & \Th{Pubmed} & \Th{NDC-Class} & \Th{Email-Enron} & \Th{Human-Disease} & \Th{Plant-Pollinator} \\
    \midrule
    HGNN(SNS) & 0.9361 & 0.8955 & \underline{0.9506}  & \underline{0.9878} & 0.9708 & 0.9554 & 0.9916 \\
    HGNN(MNS@V) & \underline{0.9558} & 0.8718 & 0.9415 & 0.9767 & 0.9702 & 0.9256 & \underline{0.9921} \\
    HGNN(CNS@V) & 0.9239 & 0.8593 & 0.9313 & 0.9767 & \underline{0.9727} & 0.9603 & 0.9917 \\
    HGNN(MNS@T) & 0.8708 & \underline{0.9353} & 0.7724 & 0.7584 & 0.9810 & 0.9569 & \underline{0.9636} \\
    HGNN(CNS@T) & 0.6955 & 0.5338 & 0.4207& 0.7772 & 0.8488 & 0.5462 & 0.6357 \\
    HGNN(HNS) & \textbf{0.9577} & \textbf{0.9623} & \textbf{0.9649} & \textbf{0.9902} & \textbf{0.9781} & \textbf{0.9663} & \textbf{0.9925} \\
    \midrule
    HNHN(SNS) & 0.9555 & \underline{0.9298} & 0.9366 &  \underline{0.9942} & \underline{0.9774} & 0.9355 & \textbf{0.9747} \\
    HNHN(MNS@V) & \underline{0.9600} & 0.9243 & \underline{0.9420} & 0.9854 & 0.9739 & \underline{0.9458} & \underline{0.9738} \\
    HNHN(CNS@V) & 0.9443 & 0.8660 & 0.8126 & 0.9851 & 0.9720 & 0.9281 & 0.8895 \\
    HNHN(MNS@T) & 0.9160 & 0.8931 & 0.8645 & 0.9346 & 0.9629 & \textbf{0.9628} & 0.9564 \\
    HNHN(CNS@T) & 0.7480 & 0.4699 & 0.3654 & 0.9274 & 0.8942 & 0.4341 & 0.8219 \\
    HNHN(HNS) & \textbf{0.9633} & \textbf{0.9459} & \textbf{0.9472} & \textbf{0.9949} & \textbf{0.9817} & 0.9422 & 0.9724 \\
    \midrule
    NHP(SNS) & 0.9653 & 0.9156 & \underline{0.9469} & \underline{0.9975} & 0.9763 & \textbf{0.9416} & \underline{0.9749} \\
    NHP(MNS@V) & \underline{0.9677} & 0.9274 & 0.9258 & 0.9952 & \underline{0.9771} & 0.9036 & \textbf{0.9756} \\
    NHP(CNS@V) & 0.9483 & 0.9048 & 0.9126  & 0.9886 & 0.9765 & 0.9136 & 0.7311 \\
    NHP(MNS@T) & 0.8993 & \underline{0.9355} & 0.4925 & 0.9728 & 0.9666 & 0.9399 & 0.8716 \\
    NHP(CNS@T) & 0.3185 & 0.3329 & 0.3880 & 0.7190 & 0.7191 & 0.3446 & 0.5868 \\
    NHP(HNS) & \textbf{0.9721} & \textbf{0.9467} & \textbf{0.9473}  & \textbf{0.9978} & \textbf{0.9817} & \underline{0.9410} & 0.9747 \\
    \midrule
    HGNNP(SNS) & \underline{0.9748} & 0.9146 & 0.9340 & \underline{0.9966} & \underline{0.9848} & \underline{0.9321} & \textbf{0.9769} \\
    HGNNP(MNS@V) & 0.9623 & \underline{0.9205} & \underline{0.9390}  & 0.9916 & 0.9843 & 0.8765 & \underline{0.9743} \\
    HGNNP(CNS@V) & 0.9586 & 0.8938 & 0.9208 & 0.9872 & 0.9845 & 0.8718 & 0.8629 \\
    HGNNP(MNS@T) & 0.9368 & 0.8816 & 0.8751 & 0.9682 & 0.9690 & \textbf{0.9586} & 0.9608 \\
    HGNNP(CNS@T) & 0.8108 & 0.4276 & 0.3487 & 0.9658 & 0.8848 & 0.3689 & 0.8257 \\
    HGNNP(HNS) & \textbf{0.9855} & \textbf{0.9456} & \textbf{0.9409} & \textbf{0.9967} & \textbf{0.9855} & 0.9309 & 0.9734 \\
    \bottomrule
    \end{tabularx}
    \caption{\textbf{Performance comparison of hypergraph negative sampling strategies across multiple datasets and model architectures, subject to AUPR}. The best and second-best results are highlighted in \textbf{black bold} and \underline{underlined}, respectively.}
    \label{tab_ap_performance}
    \end{table}
    \begin{table}[ht]
    \centering
    \scriptsize
    \setlength{\tabcolsep}{4.5pt}
    \begin{tabularx}{\linewidth}{l*{7}{>{\centering\arraybackslash}X}}
    \toprule
    \Th{Model/Data} & \Th{CiteSeer} & \Th{Cora} & \Th{Pubmed} & \Th{NDC-Class} & \Th{Email-Enron} & \Th{Human-Disease} & \Th{Plant-Pollinator} \\
    \midrule
    HGNN & 0.993381 & 0.984322 & \underline{0.993506} & \underline{0.998183} & 0.995255 & 0.988728 & 0.998732 \\
    HGNN(MNS@V) & \underline{0.993546} & 0.981088 & 0.988324 & 0.996459 & 0.995155 & 0.982109 & \underline{0.998810} \\
    HGNN(CNS@V) & 0.990172 & 0.973998 & 0.988163 & 0.996459 & \underline{0.995662} & 0.991706 & 0.998751 \\
    HGNN(MNS@T) & 0.977982 & \textbf{0.989144} & 0.962949 & 0.997098 & 0.992750 & \underline{0.992099} & 0.998368 \\
    HGNN(CNS@T) & 0.947312 & 0.897898 & 0.867618 & 0.953742 & 0.974394 & 0.828020 & 0.876932 \\
    HGNN(HNS) & \textbf{0.994288} & \underline{0.985681} & \textbf{0.993879} & \textbf{0.998507} & \textbf{0.996594} & \textbf{0.992357} & \textbf{0.998884} \\
    \midrule
    HNHN & \underline{0.994992} & 0.983027 & \textbf{0.992289} & \underline{0.999139} & \underline{0.996600} & 0.975534 & \textbf{0.995695} \\
    HNHN(MNS@V) & 0.994122 & \textbf{0.987618} & \underline{0.992269} & 0.997818 & 0.996130 & \underline{0.986982} & \underline{0.995630} \\
    HNHN(CNS@V) & 0.986838 & 0.976114 & 0.982045 & 0.997782 & 0.995833 & 0.969119 & 0.980542 \\
    HNHN(MNS@T) & 0.986812 & 0.979409 & 0.982132 & 0.989454 & 0.994304 & \textbf{0.992040} & 0.992711 \\
    HNHN(CNS@T) & 0.966680 & 0.830281 & 0.837277 & 0.988183 & 0.982767 & 0.715290 & 0.962297 \\
    HNHN(HNS) & \textbf{0.995191} & \underline{0.985147} & 0.992241 & \textbf{0.999235} & \textbf{0.997243} & 0.980612 & 0.994595 \\
    \midrule
    NHP & \underline{0.995185} & 0.988283 & \underline{0.992278} & \underline{0.999640} & 0.996327 & \underline{0.979775} & \underline{0.996146} \\
    NHP(MNS@V) & 0.995185 & 0.989081 & 0.989527 & 0.999302 & \underline{0.996481} & 0.966290 & \textbf{0.996242} \\
    NHP(CNS@V) & 0.991357 & 0.984486 & 0.986443 & 0.998303 & 0.996372 & 0.973072 & 0.944069 \\
    NHP(MNS@T) & 0.983254 & \underline{0.989849} & 0.887630 & 0.995803 & 0.994797 & \textbf{0.986991} & 0.968961 \\
    NHP(CNS@T) & 0.769260 & 0.755913 & 0.852096 & 0.917166 & 0.919789 & 0.640560 & 0.888956 \\
    NHP(HNS) & \textbf{0.995504} & \textbf{0.990547} & \textbf{0.993197} & \textbf{0.999687} & \textbf{0.997263} & 0.976132 & 0.996074 \\
    \midrule
    HGNNP & \underline{0.995138} & 0.985837 & 0.991448 & \underline{0.999511} & \textbf{0.997763} & 0.973502 & \textbf{0.996258} \\
    HGNNP(MNS@V) & 0.993309 & \underline{0.987086} & \underline{0.991871} & 0.998768 & 0.997686 & 0.961408 & \underline{0.995534} \\
    HGNNP(CNS@V) & 0.991538 & 0.985467 & 0.987300 & 0.998110 & 0.997717 & 0.960091 & 0.975209 \\
    HGNNP(MNS@T) & 0.988355 & 0.980969 & 0.982363 & 0.995127 & 0.995342 & \textbf{0.991332} & 0.993567 \\
    HGNNP(CNS@T) & 0.960648 & 0.813327 & 0.836595 & 0.994729 & 0.980405 & 0.649688 & 0.964830 \\
    HGNNP(HNS) & \textbf{0.995484} & \textbf{0.988446} & \textbf{0.992024} & \textbf{0.999528} & \underline{0.997735} & \underline{0.973707} & 0.995530 \\
    \bottomrule
    \end{tabularx}
    \caption{\textbf{Performance comparison of hypergraph negative sampling strategies across multiple datasets and model architectures, subject to NDCG}. The best and second-best results are highlighted in \textbf{black bold} and \underline{underlined}, respectively.}
    \label{tab_ndcg_performance}
\end{table}
    \begin{table}[ht]
    \centering
    \scriptsize
    \setlength{\tabcolsep}{4.5pt}
    \begin{tabularx}{\linewidth}{l*{7}{>{\centering\arraybackslash}X}}
    \toprule
    \Th{Model/Data} & \Th{CiteSeer} & \Th{Cora} & \Th{Pubmed} & \Th{NDC-Class} & \Th{Email-Enron} & \Th{Human-Disease} & \Th{Plant-Pollinator} \\
    \midrule
    HGNN & 0.027216 & 0.019543 & \underline{0.004938} & \underline{0.028049} & 0.020211 & 0.121530 & 0.029838 \\
    HGNN(MNS@V) & \underline{0.027226} & 0.019459 & 0.004903 & 0.027990 & 0.020208 & 0.120434 & \underline{0.029841} \\
    HGNN(CNS@V) & 0.027058 & 0.019225 & 0.004864 & 0.027990 & \underline{0.020233} & \underline{0.122462} & 0.029839 \\
    HGNN(MNS@T) & 0.026506 & \textbf{0.019712} & 0.004636 & 0.028014 & 0.020088 & \textbf{0.122482} & 0.029823 \\
    HGNN(CNS@T) & 0.025139 & 0.015323 & 0.002200 & 0.025604 & 0.019627 & 0.083865 & 0.016811 \\
    HGNN(HNS) & \textbf{0.027245} & \underline{0.019591} & \textbf{0.004944} & \textbf{0.028062} & \textbf{0.020268} & 0.122424 & \textbf{0.029844} \\
    \midrule
    HNHN & \underline{0.027293} & \underline{0.019497} & \underline{0.004927} & \underline{0.028083} & \underline{0.020278} & 0.117189 & \underline{0.029675} \\
    HNHN(MNS@V) & 0.027271 & \textbf{0.019713} & \textbf{0.004940} & 0.028037 & 0.020275 & \underline{0.121322} & \textbf{0.029683} \\
    HNHN(CNS@V) & 0.026779 & 0.019114 & 0.004789 & 0.028035 & 0.020265 & 0.115037 & 0.029000 \\
    HNHN(MNS@T) & 0.027047 & 0.019291 & 0.004883 & 0.027754 & 0.020213 & \textbf{0.122486} & 0.029555 \\
    HNHN(CNS@T) & 0.025919 & 0.008246 & 0.001459 & 0.027697 & 0.019896 & 0.049550 & 0.027389 \\
    HNHN(HNS) & \textbf{0.027296} & 0.019227 & 0.004896 & \textbf{0.028086} & \textbf{0.020295} & 0.118928 & 0.029529 \\
    \midrule
    NHP & \underline{0.027302} & 0.019699 & \underline{0.004930} & \underline{0.028099} & 0.020260 & \underline{0.118579} & \underline{0.029741} \\
    NHP(MNS@V) & 0.027302 & 0.019691 & 0.004864 & 0.028087 & \underline{0.020267} & 0.115082 & \textbf{0.029743} \\
    NHP(CNS@V) & 0.027121 & 0.019544 & 0.004766 & 0.028051 & 0.020265 & 0.117352 & 0.026804 \\
    NHP(MNS@T) & 0.026901 & \textbf{0.019744} & 0.002492 & 0.027954 & 0.020218 & \textbf{0.121610} & 0.027058 \\
    NHP(CNS@T) & 0.008287 & 0.003470 & 0.001801 & 0.022154 & 0.013763 & 0.029357 & 0.020933 \\
    NHP(HNS) & \textbf{0.027313} & \underline{0.019734} & \textbf{0.004941} & \textbf{0.028101} & \textbf{0.020296} & 0.117100 & 0.029737 \\
    \midrule
    HGNNP & \underline{0.027301} & 0.019512 & 0.004863 & \textbf{0.028095} & \textbf{0.020313} & 0.116507 & \textbf{0.029724} \\
    HGNNP(MNS@V) & 0.027247 & \textbf{0.019702} & \textbf{0.004940} & 0.028069 & 0.020311 & 0.115259 & 0.029659 \\
    HGNNP(CNS@V) & 0.027081 & 0.019511 & 0.004855 & 0.028046 & \underline{0.020312} & 0.114966 & 0.028731 \\
    HGNNP(MNS@T) & 0.027097 & 0.019459 & 0.004885 & 0.027944 & 0.020252 & \textbf{0.122424} & 0.029608 \\
    HGNNP(CNS@T) & 0.025708 & 0.007102 & 0.001520 & 0.027929 & 0.019738 & 0.030374 & 0.027808 \\
    HGNNP(HNS) & \textbf{0.027312} & \underline{0.019632} & \underline{0.004914} & \underline{0.028095} & 0.020298 & \underline{0.116668} & \underline{0.029677} \\
    \bottomrule
    \end{tabularx}
    \caption{\textbf{Performance comparison of hypergraph negative sampling strategies across multiple datasets and model architectures, subject to MRR}. The best and second-best results are highlighted in \textbf{black bold} and \underline{underlined}, respectively.}
    \label{tab_mrr_performance}
    \end{table}

\end{document}